\documentstyle[preprint,aps,floats,epsf]{revtex}

\newcommand{\be}{\begin{equation}}
\newcommand{\ee}{\end{equation}}
\newcommand{\bea}{\begin{eqnarray}}
\newcommand{\eea}{\end{eqnarray}}
\newcommand{\half}{{1 \over 2}}
\def\href#1#2{#2}
\def\del{\partial}
\newcommand{\tit}{{\tilde t}}
\newcommand{\tis}{{\tilde s}}
\newcommand{\tib}{{\tilde b}}
\def\IZ{\relax\ifmmode\mathchoice
{\hbox{\cmss Z\kern-.4em Z}}{\hbox{\cmss Z\kern-.4em Z}}
{\lower.9pt\hbox{\cmsss Z\kern-.4em Z}}
{\lower1.2pt\hbox{\cmsss Z\kern-.4em Z}}\else{\cmss Z\kern-.4em
Z}\fi}
\def\IR{\relax{\rm I\kern-.18em R}}
\font\cmss=cmss10 \font\cmsss=cmss10 at 7pt

\begin{document}

\preprint{IASSNS-HEP-97/113, RU-97-81, SU-ITP-97-40}

\title{Webs of (p,q) 5-branes, Five Dimensional Field Theories and Grid Diagrams}

\author{
\begin{tabular}{ccc}
{\bf Ofer Aharony} &{\bf Amihay Hanany} &{\bf Barak Kol} \\ }
\address{{\normalsize Department of Physics and Astronomy} &
School of Natural Sciences &Department of Physics\\
Rutgers University &Institute for Advanced Study &
Stanford University\\ Piscataway, NJ 08855-0849, USA  &Princeton, NJ
08540, USA &Stanford, CA 94305, USA\\ \tt{oferah@physics.rutgers.edu}
& \tt{hanany@sns.ias.edu} & \tt{barak@leland.stanford.edu}
\end{tabular}}

\maketitle
\begin{abstract}
We continue to study 5d $N=1$ supersymmetric field theories and their compactifications on a circle through brane
configurations. We develop a model, which we call $(p,q)$ Webs, which enables simple geometrical computations to
reproduce the known results, and facilitates further study. The physical concepts of field theory are transparent in
this picture, offering an interpretation for global symmetries, local symmetries, the effective (running) coupling, the
Coulomb and Higgs branches, the monopole tensions, and the mass of BPS particles. A rule for the dimension of the
Coulomb branch is found by introducing Grid Diagrams. Some known classifications of field theories are reproduced. In
addition to the study of the vacuum manifold we develop methods to determine the BPS spectrum. Some states, such as
quarks, correspond to instantons inside the 5-brane which we call strips.  In general, these may not be identified with
$(p,q)$ strings.  We describe how a strip can bend out of a 5-brane, becoming a string.  A general BPS state corresponds
to a Web of strings and strips.  For special values of the string coupling a few strips can combine and leave the
5-brane as a string.
\end{abstract}

\newpage
\begin{flushright}
To my parents, \hspace*{25pt} \\ Sara and Yoram Kol \\ BK
\end{flushright}
\tableofcontents


\section{Introduction}

The study of 5d $N=1$ supersymmetric gauge theories was pioneered by Seiberg a year ago \cite{Seiberg} using a field
theoretic approach together with results from string theory. These theories are non-renormalizable, but they may be
defined as perturbations of superconformal field theories which correspond to their strong coupling limits. A geometric
approach to the study of these theories, considering M theory compactified on a degenerate Calabi-Yau, was introduced by
Morrison - Seiberg \cite{MS} and Douglas - Katz - Vafa \cite{DKV}. The theory was further developed by Ganor - Morrison
- Seiberg \cite{GMS} and Intriligator - Morrison - Seiberg \cite{IMS}. The next subsection is a brief review of the
known results.

We continue the study of these theories using brane configurations. These were introduced in \cite{HW} for 3d theories,
and the study of the 5d case began in \cite{AH} using the $(p,q)$ 5-branes of Type IIB, and continued in \cite{BruKar}.
Brane configurations have proved successful in modeling field theories in different dimensions. The 4d $N=1$ Seiberg
dualities were described using brane configurations by Elitzur - Giveon - Kutasov \cite{EKG} and Elitzur - Giveon -
Kutasov - Rabinovici - Schwimmer \cite{EGKRS}. The 4d $N=2$ Seiberg-Witten theory was constructed via the 5M-brane (the
5-brane of M theory) by Witten \cite{Witten}, and the Seiberg-Witten curve was given a simple geometrical interpretation
(see also \cite{KLMVW}). The description of holomorphic objects in 4d $N=1$ SQCD via the 5M-brane was demonstrated by
\cite{HOO,WittenNone,BIKSY}. Many other field theory results were also derived from branes, and we will not attempt to
give a complete list of references here.

We find that in 5 dimensions the brane configurations are particularly
simple, and provide simple explanations for most of
the known field theory results (which are protected by supersymmetry, and are thus accessible using the brane
configurations).  We call the brane configuration corresponding to a five dimensional field theory, after some
rescaling, a $(p,q)$ Web. In \cite{AH} this Web was called a
``Polymeric 5-brane". We find that the geometry of the $(p,q)$ Web describes the
vacuum structure and the BPS
spectrum of the field theories. The $(p,q)$ Web reduces most of the discussion to elementary concepts, thus offering
simple derivations of known results and facilitating further study. We do not employ group theory, matrices, loop
calculations or even the classical equations of motion. In 5d $N=1$ theories, the vector multiplet includes a real
scalar. Thus, field theory symmetries (related to vector multiplets) are realized as deformation modes of the Web
(corresponding to giving expectations values to the scalars in the vector multiplets). Deformation modes with finite
mass (from the string theory point of view) in the Web correspond (in the low-energy five dimensional field theory) to
local gauge symmetries, while those with infinite mass correspond to global symmetries (associated with background
vector multiplets). For the local deformations, the effective gauge coupling is the mass of the deformation mode. The
Coulomb branch is the space of all Web configurations. The Higgs branch may be entered upon separating the Web to
sub-Webs. The prepotential can be determined via the BPS spectrum, namely the monopole tension and the BPS masses.
Monopole tensions are given by the area of faces in the Web. BPS particles are realized by Webs of strings inside the
$(p,q)$ Web, allowing a simple calculation of their masses.

The study of the $(p,q)$ Web suggests introducing certain Grid diagrams to be described in Section II. These diagrams
are dual to the Web, and seem to resemble diagrams of toric data such as in \cite{IMS,KMV}. The Grid diagram determines the dimension of the
Coulomb branch to be the number of internal points. Flop transitions of the
Grid are related to certain flops in the Web
(see section \ref{jump}). After compactifying the theory on a circle, we can lift the $(p,q)$ Web from Type IIB to M
theory. This introduces a curve \cite{BK}, which is the analogue of the Seiberg-Witten curve for a compactified 5d
theory. The relations among the triad Web - Grid - curve, and the realizations of the vacuum structure are discussed in
section II. As an application we describe the realization of the two pure $SU(2)$ theories, and the different pure
$SU(N_c)$ theories (a similar discussion was given in \cite{BISTY}). The compactified five dimensional gauge theories
were also studied in \cite{Nekrasov,LN}.

In section III we study the BPS spectrum. We find a realization for some of the
BPS particles, including the gauge instantons (which are particles in 5d), in
terms of string Webs. This allows us to count them and to determine their masses
and charges. We compare our results with the counting of BPS states using
geometric engineering, as described, for instance, in \cite{KLMVW,KKV}.
Instantonic strings within the 5-brane, that we call strips, are needed in order
to describe quarks and other states. We emphasize that, in general, instantons
inside 5-branes are not equivalent to strings. Whereas the tension of a $(p,q)$
string is $T_{p,q}^{string}=\sqrt{p^2+q^2}$, the strip tension is
$T_{p,q}^{strip}=1/ \sqrt{p^2+q^2}$ (the tensions are in string units, and the
complexified Type IIB string coupling is set to the self-dual point $\tau=i$).
Furthermore, we find that strips and strings can create ``bends" where a strip
bends out of the 5-brane to become a string.

There are various directions for further research. We described 5d gauge
theories in terms of brane configurations. There is an alternative description
in terms of M theory compactified on a Calabi-Yau manifold, with shrinking
cycles used to decouple gravity \cite{MS,DKV}. We would expect that there is a
mapping from our concepts to the Calabi-Yau language. What is the mapping
between $(p,q)$ Webs and Calabi-Yau manifolds? Our Grid diagrams seem to be
related to toric data diagrams.  In what way?
\footnote{After the appearance of this paper the relation between brane configurations and
toric geometry was discussed in \cite{LV} (see also \cite{GGM}).}

In the determination of the BPS spectrum we still miss two important points. How could it be determined whether a state
is a hypermultiplet or a vector multiplet? Which of the marginally stable states exist?

Although we reproduce most of the field theory results, there are
results that we were not able to get so far. Theories with an $SU(2)$
gauge group may have up to 7 flavors, as seen from a construction in
Type I$^\prime$ \cite{Seiberg}, but the Webs can account only for up
to 4 flavors\footnote{The introduction of an orientifold can lift this
critical number to 6, see \cite{BruKar}, but it has not been studied
in this paper.}.  Also, the brane configurations allow us to identify
the points of the Coulomb branch corresponding to roots of the Higgs
branch, from which the Higgs branch emanates, but we cannot identify
the Higgs branch itself (nor all the roots of the Higgs branch). We
can, however, describe deformations of the field theory which cause it
to go into the Higgs branch.

Most of the gauge theory examples we give will be of an $SU(2)$ gauge theory, though the discussion may easily be
generalized to $SU(N_c)$ gauge groups. The discussion could also be generalized to other gauge groups and other matter
representations using orientifold planes. See, for example, \cite{BruKar,BISTY} for $SO$ and $Sp$ groups, and
\cite{EGKRS,LL} for matter in the symmetric and anti-symmetric representations.

\subsection{A Review of Field Theory Results}

In five dimensions the minimal $N=1$ supersymmetry has 8 supercharges and the R-symmetry is $Sp(1) \simeq SU(2)$. The
small representations of the supersymmetry algebra are the vector multiplet, containing a vector field, a real scalar
and fermions, and the hypermultiplet, containing four real scalars and
fermions.

Consider a general gauge theory with a
vector multiplet in the adjoint of the gauge group $G$, and matter hypermultiplets in representations ${\bf r}_f$, with
masses $m_f$. Mass parameters for hypermultiplets in five dimensions are real, and may be viewed as background vector
multiplets. Let $g_0$ be the bare coupling of the gauge theory and denote the
instanton mass by $m_0 = 1/{g_0}^2$ reflecting the dimension of the
five dimensional gauge
coupling. The name ``instanton mass'' will be explained in the next few paragraphs.
Along the Coulomb branch, where the scalars in the
vector multiplets obtain expectation values, the non-Abelian gauge group $G$ is broken to $U(1)^r$, with $r={\rm
rank}(G)$. The theory there is described by an Abelian low-energy effective theory for the vectors ${\cal A} ^i$ in
${\cal A} =\sum_{i=1}^r{\cal A} ^iT_i$, with $T_i$ the Cartan generators of $G$. The low-energy theory is determined by
the prepotential ${\cal F}({\cal A}^i)$, which is required to be at most cubic due to 5d gauge invariance
\cite{Seiberg}. The exact quantum prepotential is given by \cite{IMS}
\be
{\cal F}=\half m_0h_{ij}\phi ^i\phi ^j + {c_{cl}\over 6} d_{ijk}\phi ^i\phi ^j\phi ^k+ {1\over 12}\left(\sum
_{{\bf R}}|{\bf R}
\cdot {\bf\phi} |^3-\sum _{f}\sum _{{\bf w}\in {\bf W}_f}|{\bf w}\cdot
{\bf \phi}+m_f|^3\right),
\label{calF}
\ee
where $\phi ^i$ are the scalar components of ${\cal A}^i$,
$h_{ij}=Tr (T_iT_j)$,  $d_{abc}=\half Tr (T_a (T_bT_c +
T_cT_b))$, ${\bf R}$ are the roots of $G$, and ${\bf W}_f$ are the
weights of $G$ in the representation ${\bf r}_f$.
$c_{cl}$ is a quantized parameter of the theory, related to a
five dimensional Chern-Simons term. In terms of
${\cal F}$ the effective gauge coupling is

\be
2 m_g (\phi)_{ij}= {\left( 1 \over g^2 \right)}_{eff \hspace{4pt} ij} = {\partial ^2 {\cal F} \over \partial \phi ^i
\partial
\phi ^j}.
\label {mgFT}
\ee

Specializing to $G=SU(N_c)$ gauge group with $N_f$ hypermultiplets in the fundamental representation of $G$, the Coulomb
branch of the moduli space is given by $\phi={\rm diag}(a_1, \dots, a_{N_c})$ with $\sum_ia_i=0$, modulo the Weyl group
action, which permutes the $a_i$, and allows us to order $a_1\ge a_2 \ge \dots \ge a_N$. The prepotential in this case,
taking $m_0 = 0$, is explicitly given by \cite{IMS}

\be
{\cal F}={1\over 12}\left(2\sum ^N_{i<j}(a_i-a_j)^3+
2c_{cl}\sum _{i=1}^Na_i^3-N_f \sum _{i=1}^N|a_i|^3\right).
\ee
The conditions on $c_{cl}$ \cite{IMS} are
\bea
c_{cl} + \half N_f \in \IZ \\ N_f + 2 \left| c_{cl}\right|\leq 2N_c
\label{c-cl2}
\eea
where the inequality (\ref{c-cl2}) is a necessary condition to have a non-trivial fixed point (which one can use to
define the gauge theory). The case $G=SU(2)$ is somewhat
special. There are two pure gauge theories labeled by a $\IZ _2$ valued theta
angle, since $ \pi _4 (SU(2)) = \IZ _2$; $c_{cl}$ is irrelevant since $d_{ijk}=0$, and the number of allowed flavors is
extended to $N_f \leq 7$.

The BPS spectrum includes electrically charged particles, instantons
and monopoles \cite{Seiberg}. By instantons we mean particles that
carry an instanton number, $I$, which is the charge under the global
$U(1)_I$ symmetry associated with the conserved current $j=*{\rm
tr}(F\wedge F)$.  In general such particles can carry gauge charges as
well. Such a charge may arise from the cubic term in the prepotential,
which contributes to the Lagrangian a term of the form $A \wedge F
\wedge F$ that couples the global instanton current to the gauge
potential. The central charge is a linear combination of all the local
and global $U(1)$ charges. For the case of a single global charge $I$,
and a single local charge $n_e$, it is given by \cite{Seiberg}
\be
Z_e = (n_e+cI) \phi + I m_0,
\ee
where $c$ is some constant related to the coefficient of the cubic terms (from here on we will absorb this coefficient
in the definition of $n_e$). The masses of BPS saturated states are equal (up to a multiplicative constant which we will
ignore here) to their central charge. Magnetic monopoles in 5d gauge theories are strings, with tensions \cite{Seiberg}
\bea
T_m \sim Z_m &=& (n_m)_i \phi_{Di}, \nonumber \\
\phi_{Di} &=& {\partial {\cal F} \over \partial \phi_i}.
\label{TmFT}
\eea

\section{$(p,q)$ Webs, Grid diagrams and Curves}

\begin{flushright}
{\it A threefold cord \\
 is not quickly broken.}\\
  (Ecclesiastes, 4, 12)
\end{flushright}

\label{general}

\subsection{Brane Configurations}

Brane configurations for 5d $N=1$ theories were constructed in
\cite{AH}, following the general method introduced in \cite{HW}.  The
configuration consists of $(p,q)$ 5-branes of Type IIB string
theory. Denote the complex scalar of Type IIB string theory by
\be
\tau = \chi/2\pi + i/\lambda,
\label{tau}
\ee
where $\lambda$ is the string coupling and $\chi$ is the axion (the RR
scalar). The tension of a $(p,q)$ 5-brane is
\be
T_{p,q} = \left| p+ \tau q\right| T_{D5},
\label{t_pq}
\ee
where $T_{D5}$ is the D5-brane tension. In our notation the $(1,0)$ 5-brane is the D5-brane and the $(0,1)$ 5-brane is
the NS5-brane. The essential geometry takes place in a plane parametrized by two real coordinates $(x,y)$, where each
5-brane is represented by a line. Four other dimensions (as well as the time dimension) are common to all 5-branes, and
provide the space-time for the field theory. The last three dimensions of the
Type IIB string theory are not used for
the description of the Coulomb branch, and correspond to deformations associated with the Higgs branch, which will be
discussed in section \ref{Higgs}. The 5-branes are permitted to form vertices provided that the \underline{$(p,q)$
charge is conserved} :
\be {\sum_{i}^{}{p_i}} = {\sum_{i}^{}{q_i}} = 0.
\label{vcharge}
\ee
Note that the $(p,q)$ label has an overall sign ambiguity which is
resolved once we choose an orientation, say going into the vertex. It
was found in \cite {AH} that a quarter of the SUSYs, which is the
required SUSY for 5d $N=1$, can be preserved provided that any $(p,q)$
5-brane is constrained to have \underline{a slope} in the $(x,y)$
plane
\be
\Delta x + i \Delta y \hspace{1pt} \parallel \hspace{1pt} p + \tau q.
\label{slope}
\ee
The last three conditions (\ref{t_pq}),(\ref{vcharge}),(\ref{slope})
guarantee the \underline{zero force condition} for vertex
equilibrium. The presence of the branes breaks the spacetime Lorentz
symmetry from $SO(1,9)$ to $SO(1,4)\times SO(3)$. The first factor is
identified with the five dimensional Lorentz symmetry, while the
double cover of the second factor is identified with the five
dimensional R-symmetry.

\begin{figure}
\centerline{\epsfxsize=160mm\epsfbox{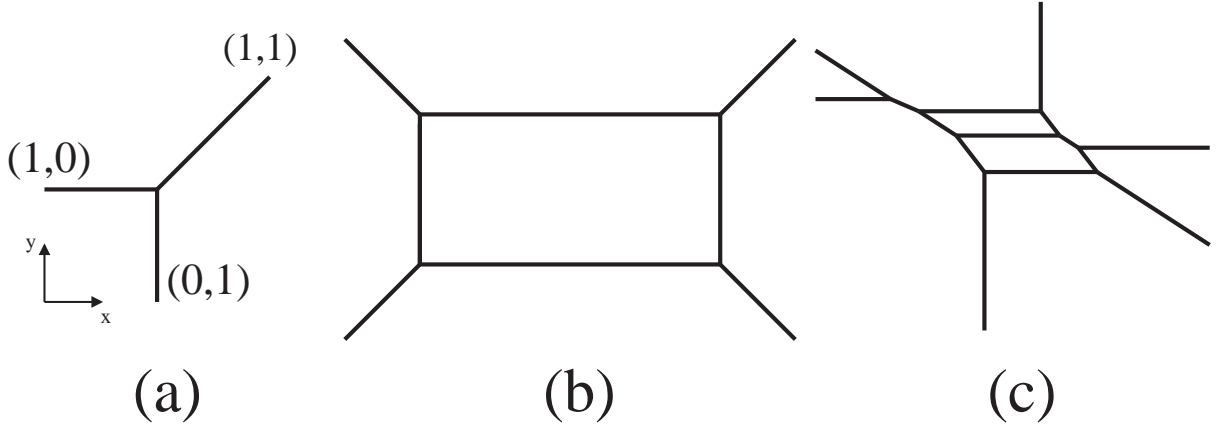}}
\medskip
\caption{Three basic brane configurations : (a) a vertex,
(b) pure $SU(2)$, (c) $N_c=3$, $N_f=2$ SQCD. Henceforth, all figures will be drawn for $\tau=i$.}
\label{basic}
\end{figure}

Three examples are shown in figure \ref{basic} (all figures will be
drawn for $\tau=i$).  Figure \ref{basic}a is the simplest
non-trivial configuration, which is a vertex of $(1,0)$, $(0,1)$ and
$(1,1)$ 5-branes. Figure \ref{basic}b shows the simplest
configuration of a gauge theory, which is the pure $SU(2)$ case.  When
$N_c$ finite parallel branes can be made to overlap one gets an
$SU(N_c)$ gauge group \cite{WittenSUN}. Figure \ref{basic}c
demonstrates how to add quarks to the configuration with horizontal
semi-infinite branes \cite{HW}, resulting here in a $N_c=3, N_f=2$
SQCD theory.

The usual \underline{conditions required to get a low energy 5d field theory} on the brane are that gravity decouples,
namely
\be
E \ll M_p,
\label{gravcond}
\ee
where $E$ is the energy scale on the brane and $M_p$ is the Planck mass, and
that the massive Kaluza Klein modes can be
integrated out
\be
E \ll 1/\Delta x, 1/\Delta y,
\label{kkcond}
\ee
where $\Delta x, \Delta y$ are the largest length scales in the configuration.
We also require that the low energy 5d theory will decouple from the 6d theory
on the semi-infinite branes\footnote{Indeed, the massless modes in 5d will be
seen not to have any component transverse to an external leg.}. In general, this
low energy theory allows us to study both the vacuum manifold and the 5d BPS
states.  \underline{Parallel external legs} present a problem. Strings
stretching between them are states of the 6d theory that are charged under the
global symmetry. This might cause the global charge to be non-conserved in the
5d theory in processes at energies above the mass of these states, thus further
limiting the energy range. Configurations with parallel external legs sometimes
also lead to directions in moduli space where the superpotential is not strictly
convex, and a strong-coupling fixed point theory is not obviously well-defined.
This could be interpreted to mean that such a configuration is consistent as a
sub - diagram, and could become well defined in the UV after being embedded in a
larger configuration (namely, we can flow to these theories from other theories
with a well-defined fixed point).

\begin{figure}
\centerline{\epsfxsize=100mm\epsfbox{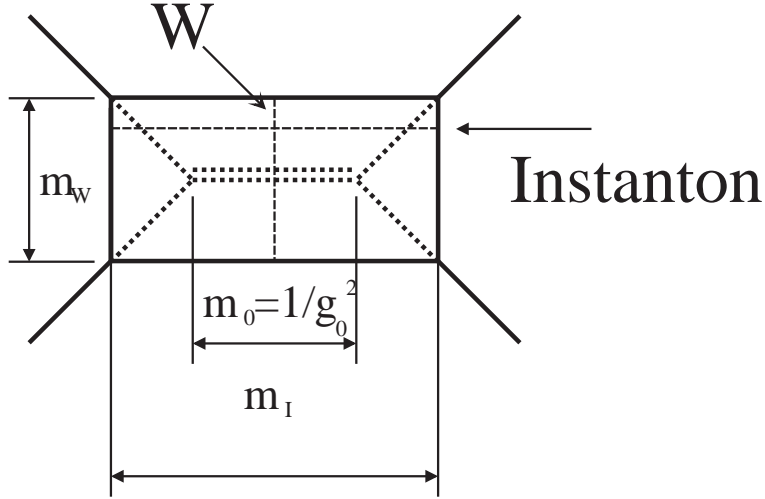}}
\medskip
\caption{The basic BPS states in the pure $SU(2)$ gauge theory --
the W boson and the instanton. The solid lines correspond to a generic
brane configuration on the Coulomb branch of this theory, while the
dotted lines correspond to the origin of the Coulomb branch.}
\label{su2_BPS}
\end{figure}

In the case of the pure $SU(2)$ gauge theory there are two field theory parameters that can be read off from the brane
configuration (figure \ref{su2_BPS}). A fundamental string stretched between the horizontal D5-branes is known to be BPS
saturated and to correspond to the W boson. Its mass is
\be
m_W = \Delta y T_s,
\label{mw}
\ee
where $T_s$ is the fundamental string tension. $\Delta y$ is
proportional to the scalar $\phi$ in the low-energy $U(1)$ vector
multiplet, which we will normalize so that $m_W = \phi$.
Similarly, we may consider a
D-string stretched between the vertical NS
branes. It is BPS saturated and we will argue below that it
is an instanton.
For $\chi=0$ its mass is given by
\be
m_I = \Delta x  \left|\tau\right|T_s.
\label{mi}
\ee

\subsection{$(p,q)$ Webs}
\label{Webs}

Define an abstract \underline{$(p,q)$ Web} to correspond to a brane
configuration as described above, with
\be
\tau=i.
\ee
Namely, it consists of line segments in a (rescaled) plane $(x,y)$, each labeled by relatively prime integers $(p,q)$. A
segment is constrained to have a slope
\be
\Delta x : \Delta y = p : q,
\ee
and vertices conserve $(p,q)$ charge.

Without loss of generality we can always ``normalize" $\tau$ in this
way. $\tau$ plays the role of a \underline{redundant parameter} as far
as the low energy field theory is concerned. Take as an example the
pure $SU(2)$ gauge theory, figure \ref{basic}b.
When the axion vanishes, we can compensate
for a change in the string coupling by rescaling $x$ and $y$, keeping the
physical field theory parameters $m_W$ and $m_I$ (\ref{mw}),(\ref{mi})
fixed. Similarly, we can compensate for a non-zero axion with a linear
transformation that tilts the plane while keeping $\Delta y$
fixed. The general rule is that the Web should remain fixed in the
normalized coordinates $[\tilde{x},\tilde{y}]$, which are coordinates
taken in the $[1,\tau]$ basis
\be
\left[\matrix{
x\cr y\cr }\right]
= \tilde{x}
\left[\matrix{
1\cr 0\cr }\right] + \tilde{y}
\left[\matrix{
Re(\tau)\cr Im(\tau)\cr }\right].
\label{norm_xy}
\ee
We conjecture that when $\tau$ and the configuration are changed in this manner, the low-energy field theory remains
unchanged. Thus, the $\tau$ parameter is redundant in the same way that the Type IIA string coupling constant was found
to be redundant for 4d configurations \cite{Witten}. Since $\tau$ is defined up to an $SL(2,{\IZ })$ transformation,
there is a
\underline{residual $SL(2,\IZ )$} operating on the plane :
\bea
\left[\matrix{
x'\cr y'\cr }\right]
= \text{\LARGE A}
\left[\matrix{
x\cr y\cr }\right] \nonumber \\ A \in SL(2,\IZ ).
\label{web_resid}
\eea
Note that the valid energy range for the low energy field theory
(\ref{gravcond}),(\ref{kkcond}) can change under these
transformations and rescaling.

Distances in the plane of the Web can be converted to mass units using the
string tension. That turns out to be
convenient in describing a fixed field theory while taking the Type IIB limit of
M theory (section \ref{curves}), since it
fixes the masses of BPS states.

\begin{figure}
\centerline{\epsfxsize=100mm\epsfbox{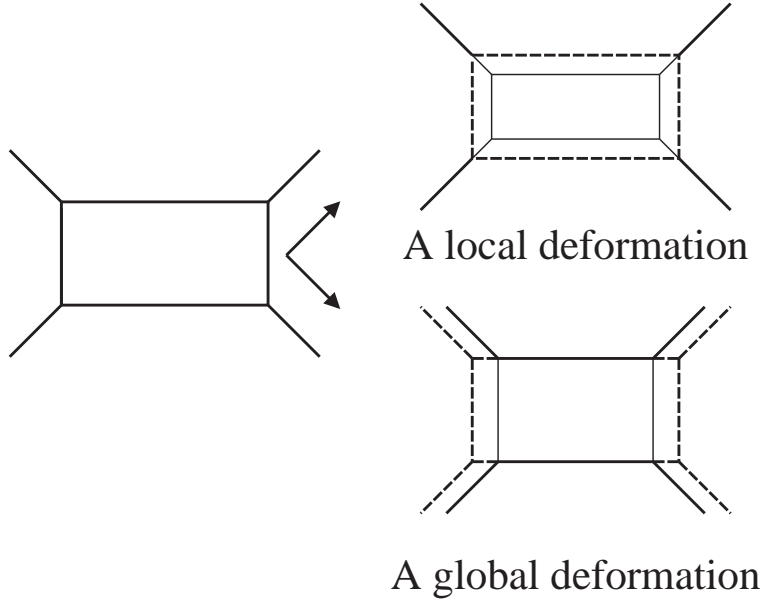}}
\medskip
\caption{Deformations of a $(p,q)$ Web. Dashed lines denote the old
configuration.}
\label{deform}
\end{figure}

Consider the allowed deformations of a $(p,q)$ Web. An \underline{external leg} is an edge that is semi-infinite. During
a deformation of a Web the number of external legs and their labels do not change. An example of two kinds of
deformations of the pure $SU(2)$ theory is given in figure \ref{deform}.
The equilibrium condition for the vertices guarantees that these deformations
do not cost energy -- they are zero modes.
 The location of external legs defines the \underline{asymptotic configuration}. A
\underline{local deformation} is a deformation that does not change the
asymptotic configuration, and thus the moving edges have finite mass in the Web
(by this we really mean a finite mass density in the $(x,y)$ plane; of
course the mass of any edge is really infinite because of the infinite
5d spacetime). Thus, it can be described in terms of VEVs of fields in
the 5d field theory. In the brane constructions all such deformations
correspond to giving VEVs to scalars in vector multiplets, associated
(at generic points in the moduli space) with local $U(1)$ gauge
fields. A \underline{global deformation} is one that does move the
external legs, and thus its mass is infinite in the Web
(since it is proportional to the length of the external legs). Such
deformations are changes of parameters from the point of view of the
5d field theory, and are associated with global $U(1)$ charges (the
positions of the asymptotic 5-branes are associated with scalars
in background vector multiplets).

The mass of the local deformation mode determines the
\underline{effective gauge coupling}, which is the metric on the
Coulomb branch moduli space in the 5d gauge theory. In the brane configuration we can determine this by the kinetic
energy of the local deformation in the string theory. Take a local deformation mode corresponding to moving the edges
around a particular face in the Web\footnote{Note that the fact that such a deformation involves moving all the 5-branes
surrounding a face in the Web implies that the low-energy $U(1)$ gauge field is a combination of the gauge fields from
all the 5-branes surrounding a face in the Web, and not just from the D5-branes as one might have naively expected.}. In
the field theory, this is associated with a scalar $\phi$ (in a $U(1)$ vector multiplet), whose kinetic term is (in our
normalization) $(1/4g_{eff}^2)(\phi) \del_\mu \phi \del^\mu \phi$. In the string theory, changing $\phi$ to $\phi+\delta
\phi$ moves the $i$'th edge surrounding the face by a distance $\delta_i \delta \phi / T_s$ for some constant
$\delta_i$. If the mass (density) of this edge is $m_i$, the corresponding contribution to the kinetic term in the
string theory will then be proportional to $\half m_i (\delta_i \delta
\phi)^2$. Thus, we find that the metric on the Coulomb branch is given by the sum of the masses of the edges which
are moved by the deformation, weighted by their displacements squared :
\be
{\left( 1 \over g^2 \right)}_{eff} = 2m_g =2(\text{weighted mass of local mode})= 2 \sum_{edges} m_i \delta _i^2,
\label{mg}
\ee
where $\delta _i$ is the vector of relative amplitudes in the mode.

Monopoles in 5d gauge theories are strings, which can be realized as a 3-brane wrapping a face in the Web \cite{AH}.
Thus, \underline{the monopole tension} is given by the area of a face, measured in units of the 3-brane tension $T_s^2$
(recall that $\tau=i$) :
\be
T_{m} = \text{Area of face.}
\label{Tm}
\ee
Since the perimeter is the differential of the area, we see that
the last two equations (\ref{mg}),(\ref{Tm}) are related
to the field theory definition of these quantities in terms
of the prepotential (\ref{TmFT}),(\ref{mgFT}).

To determine the number of global deformations, $n_G$, we note that
there is
one deformation associated with each external leg,
but not all of them produce a new Web.
Denote the number of external legs by $n_X$. We can discard two deformations
that result in translations of the whole brane configuration in the $(x,y)$
plane (which correspond to two momenta that cannot be carried by the 5d theory).
Additional deformations
are lost due to the constraints of the Web. Consider a connected component of the configuration. When we have moved all
but one of the external legs, the location of the last one is already determined. Assuming the diagram is connected this
amounts to discarding one additional deformation. So, we find
\bea
\text{rank(global group)} && =\sharp \text{(global deformations)=
$\sharp$(external legs) - 3,} \nonumber \\
n_G &&= n_X-3.
\label{no_global}
\eea
In addition we have an $SO(3)$ global symmetry associated with
rotations in the 3 unused dimensions.
This is identified with the $Sp(1)_R$ global R-symmetry.

\begin{figure}
\centerline{\epsfxsize=100mm\epsfbox{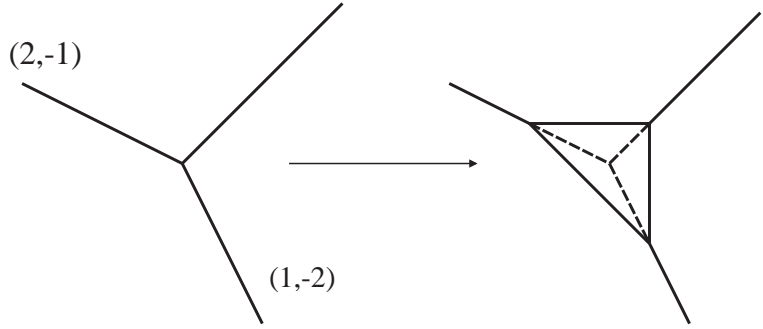}}
\medskip
\caption{A ``hidden" face in the $E_0$ Web, realizing the 1d Coulomb branch.}
\label{e0}
\end{figure}

The number of local deformations, $n_L$, is related to the number of internal faces in the diagram. Indeed, every finite
edge has one degree of freedom corresponding to its transverse position in the plane. These degrees of freedom are
constrained since the edges must meet at vertices. Each vertex contributes one constraint equation, if we assume that 3
edges intersect at each vertex (a vertex with more edges can always be decomposed into a number of 3-edge vertices).
However, these constraints are not independent. The number of relations among them equals the number of connected
components of the graph, which we will assume now to be one. Thus we find the number of local deformations
\bea
n_L= \text{rank(local gauge group)} =&& \sharp\text{(local deformations)}= \nonumber \\
 =&& E(internal) - V + 1  = F(internal)
\label{dim_coulomb_i}
\eea
where $E$ is the number of edges, $V$ is the number of vertices, $F$ is the number of faces, and we used Euler's formula
$V - E + F = 1$ for the compact part of the configuration.   Some vertices might be resolved and create new faces.
Figure \ref{e0}, called the $E_0$ theory \cite{AH}, is an example of a vertex that can be resolved to create a new
face, whereas the basic vertex (figure \ref{basic}a) cannot be resolved. In order to find the total number of local
deformations, for a given asymptotic configuration, we will need a new tool which will be developed in the next
subsection.

\underline{Example}: Consider the case of pure $SU(2)$ gauge theory
(figure \ref{su2_BPS}).
\begin{itemize}
\item{The rank of the global group is $n_G=n_X-3=1$ (\ref{no_global}),
and in this case it is just the $U(1)_I$ symmetry. The parameter associated with
this group is $m_0=1/g_0^2$.}
\item{The rank of the local group on the Coulomb branch is one
(\ref{dim_coulomb_i}), and at a generic point it is just $U(1)$.
The parameter on the Coulomb branch is $m_W = \phi$. For
$\phi=0$ the W boson is massless, and the
gauge symmetry is enhanced to $SU(2)$.}
\item{For $\phi=m_0=0$ we have a conformal theory with $E_1=SU(2)$
global symmetry \cite{Seiberg}. Enhanced global symmetries will be
discussed in section \ref{Enhanced}.}
\item{The instanton mass is $m_I=m_0+\phi$.}
\item{The effective coupling
is (\ref{mg})
\be
(1/g_{eff}^2)(\phi) =2m_g(\phi) =2\cdot ({1\over 2})^2\cdot (m_W+m_W+m_I+m_I) = m_0+2\phi,
\ee
since in this configuration $\delta_i=1/2$ for all four edges.}
\item{The monopole tension is, using (\ref{Tm}),
$T_m(\phi)=m_I \phi =(m_0+\phi)\phi$.}
\item{Let us find the superpotential ${\cal F}$. From the last two
equations we have $2 m_g=1/g_{eff}^2=\partial T_m /\partial \phi$. Thus, we find using the coordinate $\phi$ for the
Coulomb branch that
\be
{\cal F}(\phi)={1 \over 2}m_0 \phi^2 + {1 \over 3} \phi^3.
\ee}
\end{itemize}

\subsection {Grid Diagrams}

A Grid diagram is defined on a 2d integer lattice labeled by coordinates
$(a,b)$. We shall denote its components by
\underline{points, lines and polygons} to distinguish them from the
components of the Web, which we call vertices, edges and faces. The
diagram consists of points, which lie on the Grid, and of lines
joining them. The contour of the diagram is convex, and so are the
internal polygons inside it. There might be more conditions on the
diagram, but rather than state all of them, we shall describe how to
build it.

\begin{figure}
\centerline{\epsfxsize=100mm\epsfbox{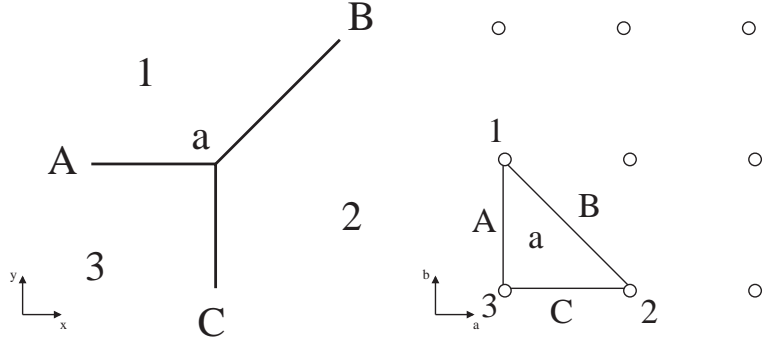}}
\medskip
\caption{The Grid diagram for the simple vertex of figure \ref{basic}a.
Vertices and corresponding polygons are marked a,b,c,..., edges and
corresponding lines are marked A,B,C,..., and faces and
corresponding points are
marked $1, 2, 3, \dots$}
\label{basic_grid}
\end{figure}

\begin{figure}
\centerline{\epsfxsize=100mm\epsfbox{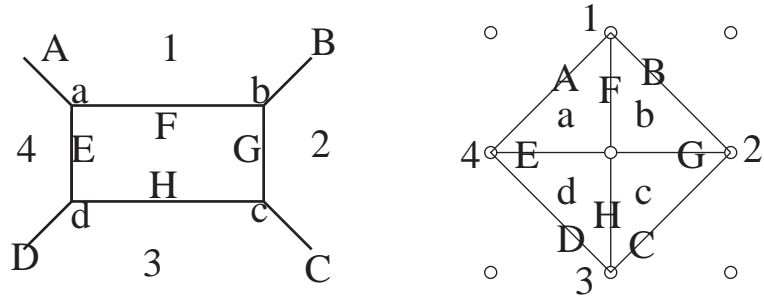}}
\medskip
\caption{The Grid diagram for the pure $SU(2)$ gauge theory
of figure \ref{basic}b.}
\label{su2_grid}
\end{figure}

\begin{figure}
\centerline{\epsfxsize=100mm\epsfbox{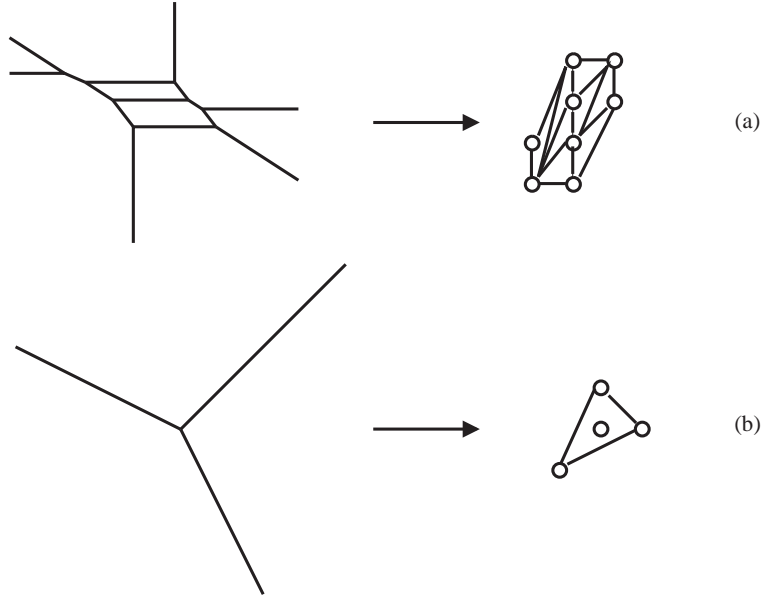}}
\medskip
\caption{Grid diagrams for (a) the $SU(3)$, $N_f=2$ SQCD theory
(figure \ref{basic}c), and
(b) the $E_0$ theory.}
\label{more_grid}
\end{figure}

Given a $(p,q)$ Web we will associate with it a Grid diagram. The Grid diagram will be the \underline{dual graph} to the
Web, exchanging vertices with polygons and faces with points. \underline{The line corresponding to a $(p,q)$ edge is
orthogonal to it} and is represented by the Grid vector $\pm (-q,p)$. The Grid diagram for the simple vertex (figure
\ref{basic}a) is shown in figure \ref{basic_grid}. Further examples of Grid diagrams for Webs that we already encountered
are given in figures \ref{su2_grid} and \ref{more_grid}.

To construct a Grid diagram one starts by marking an arbitrary vertex on the
Grid which is chosen to correspond to some
face in the Web.  Crossing to an adjacent face requires passing through a $(p,q)$ edge, and thus an orthogonal line
represented by the Grid vector $\pm (-q,p)$ should be marked, ending in a point that represents the adjacent face.
Consistency requires that if we go around a closed loop of faces in the Web, circling a vertex, we will return to the
same point in the diagram. This is guaranteed by the charge conservation property of the vertex :
\be
\sum_{\text{lines $\in$ polygon}}[p,q] =
\sum_{\text{edges $\in$ vertex}}[-q,p] = 0.
\ee
Circling infinity in the Web, we see that \underline{the Grid has to be convex}. A point that has a line going through
it without changing direction is said to be \underline{dividing the line}. Having no parallel external legs corresponds
to the Grid being ``strictly convex" -- not having an external point that divides a line.

Compare the Grid diagram for the basic vertex (figure \ref{basic_grid}) with the Grid diagram for the $E_0$ theory
(figure \ref{more_grid}b). The difference between them is that the diagram for the $E_0$ theory has an internal point.
Using a property of the dual graph, we can now compute the dimension of the Coulomb branch :
\bea
n_L &=& \text{
\underline{dim(Coulomb branch)} = $\sharp$(local deformations)} = \nonumber \\
&=& \text {$\sharp$(possible internal faces in the Web)
= \underline{$\sharp$(internal points in the diagram)}.}
\label{dim_coulomb}
\eea

The residual $SL(2,\IZ )$ symmetry that we encountered for the Web
(\ref{web_resid}) translates to an $SL(2,\IZ )$ symmetry of the Grid.

While the Grid diagram retains the information about the edges in the
Web, their slopes and vertices, it ``forgets" their sizes and
locations. Given a Grid diagram we can construct the Web diagram,
modulo the sizes of the edges, as the dual graph to the diagram,
remembering that edges are orthogonal to the lines they cross.


\subsection {Curves}
\label{curves}

In this section we review the description of curves that can be
associated to the brane configuration upon compactification on a
circle. It appeared in a preliminary report \cite{BK}, and was further
developed by Brandhuber, Itzhaki, Sonnenschein, Theisen and
Yankielowicz \cite{BISTY}.

Following \cite{Witten}, an M theory description will be sought, in order to smooth out the geometrical singularities at
vertices. Type IIB string theory itself has no non-singular M theory description, so we cannot get such a smoothed
description for the 5d theories. However, Type IIB string theory compactified on a circle of perimeter $L_B$ is dual to M
theory compactified on a torus \cite{Schwarz,Aspinwall}, with a base
of length $2\pi L_t$ and a modular parameter $\tau$.
\underline{The relations between the parameters of Type IIB and M theory are}
\bea
\tau _{IIB}=\tau \text{(torus)}&&  \qquad
T_s = 2\pi L_t T_M \nonumber \\ L_B &&= 1 /[2\pi L_t^2 Im(\tau) T_M],
\label{2BM}
\eea
where $T_M={1 \over (2\pi)^2 l_{11}^3}$ is the M theory
membrane tension, $l_{11}$ is the Planck length in 11d, and $T_s$ is the
fundamental Type IIB string tension.

A 5-brane of the Type IIB theory compactified on a circle has two possible M
theory origins. It can arise either from a
5M-brane, or from a KK monopole associated with the compactification. Following
\cite{Witten}, we wish to look at a
configuration involving a 5M-brane.  An unwrapped $(p,q)$ 5-brane in Type IIB
string theory is associated with a KK
monopole of M theory (an equivalent description was recently given in
\cite{newWitten}). However, a $(p,q)$ 5-brane
wrapped around the Type IIB circle is identified with a 5M-brane wrapping a
$(p,q)$ cycle on the torus. Thus, in M
theory a brane configuration of the sort described above, when compactified on
\underline{a compact dimension $L_4 = L_B$},
is described by a single 5M-brane \cite{BK}.

Denote the coordinates on the M theory
torus by $(x_t,y_t)$. The slope condition (\ref{slope}), when
translated into M theory, requires
that the slope in the
$(x,y)$ plane equals the one in the $(x_t,y_t)$ plane.
By introducing \underline{complex coordinates}
\be
x + i x_t  \qquad y + i y_t
\label{complex}
\ee
the \underline{slope condition is transformed into requiring analyticity}.
Indeed, in M theory, the BPS condition
translates \cite{BBS,Witten} to the ``supersymmetric cycle"
condition.
As defined above, for $\tau=i$, the complex coordinates both live on a
cylinder of periodicity $2\pi i L_t$. Define
single valued dimensionless complex coordinates

\be
s = \exp( (x+i x_t)/L_t),  \qquad t = \exp ((y +i y_t) / L_t).
\label{st}
\ee
In these coordinates the 5M-brane configuration is defined by a
surface $S$, in the ambient space $M= \IR^2
\times {\bf T^2}$ parametrized by $(x,x_t,y,y_t)$, which can be written in the form
\be
F (s,t) = 0.
\ee

\underline{The charge conservation condition is transformed into a topological
identity}. The surface $S$ has one hole for each external leg. There is one
homology cycle $c_i$ on $S$ related to each hole, satisfying that their sum is
homologically trivial. When we consider them as cycles in $M$, they are the
$(p,q)$ cycles of the external legs :
\be
\sum_{holes} c_i = 0 \text{ in } H_1(S) \Rightarrow \sum
\left[\matrix{
p\cr q\cr }\right]_i =0 \text{ in } H_1(M).
\ee

Having reviewed the basic setting for the curves, we shall now describe how to find the curve for a given $(p,q)$ Web.
Consider a $(p,q)$ edge in the Web. It is given by a linear equation of the form $m+(-qx+py)T_s=0$, where $m$
corresponds to the transverse position of the edge, and it has dimensions of mass. Generally $m$ can be either a
parameter of the field theory, such as $m_0$, or the VEV of a field (like $\phi$). Translating to the $(s,t)$ variables
we have $A s^{-q} t^p = 1$, where $\left| A\right| = \exp (m / L_t
T_s) = \exp (m L_4)$. Thus, in the
vicinity of the edge (and in the 5d limit) the curve consists of two monomials
of the form $F(s,t) \sim A_1 s^{a1}
t^{b1} + A_2 s^{a2} t^{b2}$, where
\be
(a1,b1)-(a2,b2)=(-q,p), \hspace{2pt} \left| A_1/A_2 \right|= \exp (m L_4).
\label{monom}
\ee
The integer vector $(-q,p)$ reminds us of the Grid diagram. Indeed,
we can associate a monomial $A s^a t^b$ with every point
$(a,b)$ in the Grid, and then the curve is just the sum of these
monomials ! This is the general relation between $(p,q)$ Webs and
curves anticipated in \cite{BK}.

Generally, the curve is a polynomial of the form $F(s,t)=\sum f_i, f_i(s,t)=A_i s^{a_i} t^{b_i}$.  Each monomial $f_i$
corresponds to a face in the Web where it is dominant. We have to verify that
coefficients $A_i$ can be assigned such
that the monomials condition (\ref{monom}) is satisfied. Consider a loop of $n$
points in the Grid circling a polygon.
Assign the first coefficient $A_1$ arbitrarily. The coefficients $A_2,\dots , A_n$ are determined by (\ref{monom}), and
consistency requires that the last edge will be described correctly by $A_n$ and $A_1$. This polygon in the Grid is
mapped to a vertex in the Web of coordinates $(x_0, y_0)$. The edge equations give
\be
\left|f_1(x_0,y_0) \right|= \left|f_2(x_0,y_0) \right|= ... =
\left|f_n(x_0,y_0)
\right| \Longrightarrow \left|f_n(x_0,y_0) \right|= \left|f_1(x_0,y_0) \right|
\ee
so that indeed the last edge passes through the same vertex as well. Note that
although this equation constrains only
the absolute value of $f(x_0,y_0)$, there is a freedom in choosing its phase
through the choice of $x_{t0},y_{t0}$.

\begin{figure}
\centerline{\epsfxsize=80mm\epsfbox{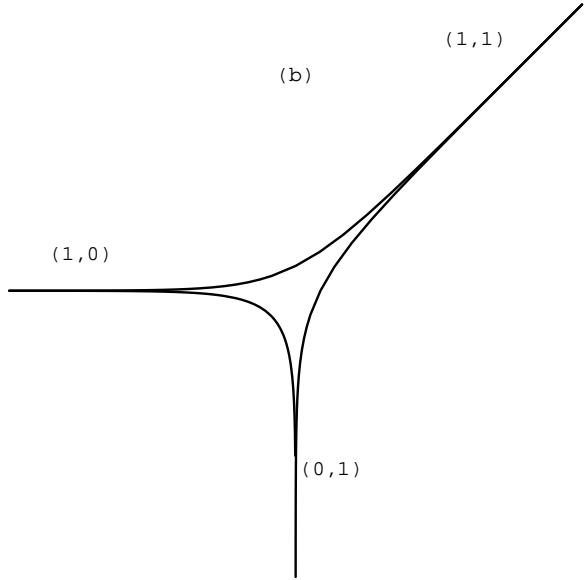}}
\medskip
\caption{A projection of the smooth curve for the simple vertex
(figure \ref{basic}a).}
\label{smooth_vertex}
\end{figure}

Turning to examples we can determine the curves for the Webs considered
previously. Using the relevant Grid for the
vertex of figure \ref{basic_grid} we get (after setting three of the
coefficients to one as described below)
\be
F(s,t) = 1 + s + t = 0.
\label{cvertex}
\ee
The projection of this curve on the $(x,y)$ plane is shown in figure
\ref{smooth_vertex}.

For the pure $SU(2)$ gauge theory of figure \ref{su2_grid}, a
convenient parametrization for the curve is given by
\be
F(s,t) = As + t + AB st + As t^2 + t s^2.
\label{su2_curve}
\ee
For large $L_4$ (corresponding to the 5d limit), projecting on the
$(x,y)$ plane and comparing with figure \ref{su2_BPS}, we find that
the relation between the parameters of the curve and of the field
theory is given by
\be
A \sim 2\exp({1 \over g_0^2} L_4/2), \hspace{2pt} B \sim 2\exp(m_W L_4/2).
\label{relation}
\ee

As we take the 5d limit $L_4 \to \infty$, \underline{the curve approaches the Web up to small corrections}. From the
curve for the vertex (\ref{cvertex}) and figure \ref{smooth_vertex} we see that around the vertex the corrections are
significant at distances of the order of
\be
\delta \text{(vertex)} \sim 1/L_4 T_s.
\ee
Away from the vertex, the scale of the corrections
decreases exponentially with the distance
$m$ (in mass units as above) from the vertex :
\be
\delta \text{(edge)} \sim \exp (-m L_4)/L_4 T_s.
\label{edge_cor}
\ee
In the 5d limit, all these corrections disappear and we recover the Web.

\underline{The curve} has a \underline{central role} in describing \underline{the 5d theory
compactified on a circle}, in analogy with the role of the
curve in 4d $N=2$ Seiberg - Witten theory \cite{SW}. The curve depends
on the parameters of the theory and its moduli, and it determines the
masses of BPS states, the metric on moduli space and the singularities
in moduli space.  It carries full non-perturbative information, as
anticipated in \cite{BK}.

\underline{Global parameters and moduli translate into the coefficients of the
curve}. We call a monomial external or internal according to the corresponding point in the Grid. An external
coefficient is a global parameter, and an internal coefficient is a modulus (a VEV of a scalar field in the theory).
Three of the coefficients $A_i$ can be eliminated (chosen to any value). One may be eliminated due to the freedom of
multiplying $F(s,t)$ by an overall factor, and two more may be eliminated by rescaling $s$ and $t$ by constants
(corresponding to translations in $(x,y,x_t,y_t)$). This agrees with our previous counting (\ref{no_global}),
(\ref{dim_coulomb}). Whereas the parameters and moduli of the Web (describing the uncompactified 5d theory) were real,
the parameters of the curve are complex. This happens because in the compactification from 5d to 4d the Wilson loop is
added as a modulus, so the scalar in the vector multiplet becomes complex, with period $2 \pi i/L_4$. The parameters of
the theory, which may be viewed as background vector multiplets, similarly become complexified.

\underline{The masses of BPS states are given by the mass of minimal area
membranes} that end on the 5M-brane \cite{Witten}
\be
dm_{BPS}^2 = d\text{Area}^2 = \left|dx dy\right|^2 +\left|dx_t dy_t\right|^2 + \left|dx dy_t\right|^2 + \left|dy
dx_t\right|^2 +
\left|dx dx_t\right|^2 + \left|dy dy_t\right|^2.
\ee
The BPS configurations are exactly the ones where we can replace the area
integration by integrating a closed two-form
\cite{Mikhailov}. (For a related discussion see also \cite{FS}.)
The integration of the two-form over the membrane can be replaced with an
integration over the
boundary, and since the integration is over a holomorphic form, the integral
depends only on the homology class of the
boundary cycle
\bea
\label{two_form}
m_{BPS} =&& \int_{membrane}^{}{{ds \over s} {dt \over t}} =
\int_{\partial membrane}^{} {\lambda,} \\ d\lambda =&& {ds \over s} {dt
\over t}.
\eea
Note that the closed two form (\ref{two_form}) always has a primitive, $\lambda$, since the membrane does not wrap the
torus. The choice of a primitive depends on the configuration. For example, for the pure $SU(2)$ configuration with a W
boson one should use $\lambda =- ds/s
\text{ }log(t)$, and for the instanton $\lambda = log(s) dt/t$. The
effective gauge coupling is determined in terms of the BPS states, as
\be
2 m_g = {1\over g_{eff}^2}
= {\partial T_m \over \partial \phi} \sim {\partial a_D \over
\partial a}.
\ee
We describe how to find the singularities in moduli space in section \ref{sing}.

A field theory analysis of the compactified 5d theory was carried out
by Nekrasov, and by Lawrence and Nekrasov \cite{Nekrasov,LN}. The
motivation there comes from integrable systems where the role of the
fundamental two form of the integrable system is played by $ds/s \cdot
dt/t$. The analysis includes a one-loop calculation, an
instanton calculation, and finding the curve that gives full
non-perturbative results. Curves are given for different $N_c, N_f$. These
results seem to match with ours under a change of variables
\bea
dp \sim {ds \over s} \qquad dq \sim {dt \over t} \\ U \sim \text{an internal parameter,}
\eea
although we were not able to show full agreement. $p,q$ are the conjugate variables of the integrable system, and $U$
parametrizes the moduli space. The geometrical edge correction (\ref{edge_cor}) can be interpreted as a world-line
instanton effect from a particle of mass $m$.

\begin{figure}
\centerline{\epsfxsize=140mm\epsfbox{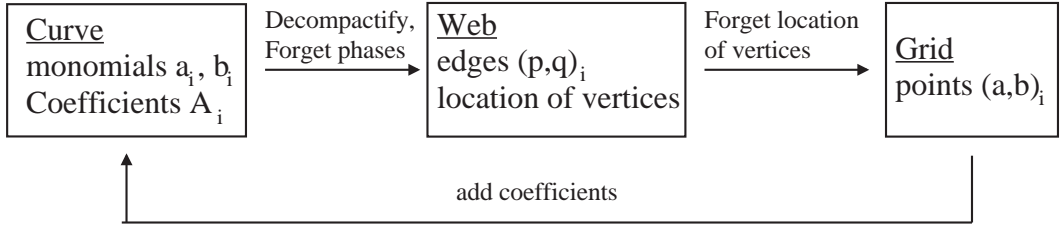}}
\medskip
\caption{The information flow between the curve, the Web and the Grid.}
\label{information}
\end{figure}

To summarize the relation between Webs, Grids and curves consider the figure of information flow (\ref{information}).
The curve carries the most information. Passing to the 5d limit, $L_4 \to \infty$, we lose all the phase information and
get the Web. The monomial data $(a_i,b_i)$ determines the edge labels $(p_i,q_i)$, and the absolute value of the
coefficients $\left| A_i\right| $ determines the locations of the edges. Passing from the Web to the Grid we lose the
information on distances in the $(x,y)$ plane, but keep the monomial data. To return from the Grid to the curve we have
to add in the coefficients of the polynomial.

\subsection{Classification of 5d Theories Using Grids}
\label{SUtwoGen}
\label{SUn}

We turn to some applications of the Grid diagrams. We rederive the possible pure $SU(2)$ gauge theories, and the
possible values of $c_{cl}$ for pure $SU(N_c)$ gauge theories found in
\cite{IMS}. We present other theories that are
easy to describe using the Grid, but have no known Lagrangian description.

\begin{figure}
\centerline{\epsfxsize=130mm\epsfbox{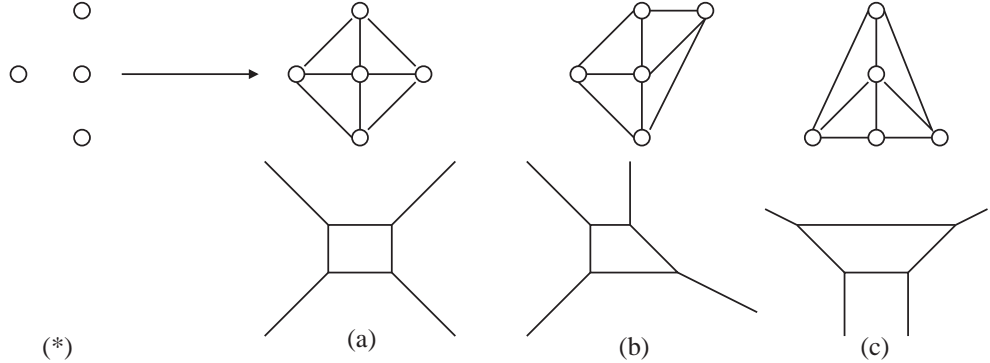}}
\medskip
\caption{The possible $SU(2)$ theories from Grids. Starting with the basic Grid
(*) we can construct both $\Theta=0$ (mod 2) in (a), and $\Theta=1$ (mod 2) in
(b). Figure (c) does not seem to define a new theory.}
\label{su2_options}
\end{figure}

Consider the possible configurations that result in \underline{a pure $SU(2)$ gauge theory}. To get a point of enhanced
$SU(2)$ we need two parallel branes, which we can choose to be horizontal.  This translates into three vertically
adjacent points in the Grid. Since we want the Coulomb branch to be one dimensional, the middle point should be the only
internal point of the diagram. Since we want to have just one global charge -- the instanton charge -- we need 4
external points (\ref{no_global}). So we have to add two external points, one to the right of the vertical column and
one to the left. Using the residual $SL(2,\IZ )$ symmetry (section \ref{Webs}) we can move the left point to the center,
as shown in the figure \ref{su2_options}*. There are 3 options for the right point consistent with convexity (up to an
obvious $y \to -y$ parity symmetry) resulting in 3 different Webs, figure \ref{su2_options}(a-c). For reasons which
will be described below, we recognize option (a) to be the $SU(2)$ gauge theory with a vanishing theta angle, $\Theta=0$
(mod 2)(in $\pi_4(SU(2)) = \IZ_2$), which is a deformation of the $E_1$ fixed point theory \cite{AH}. We recognize
option (b) to be $SU(2)$ with $\Theta=1$ (mod 2), a deformation of the $\tilde{E}_1$ fixed point theory \cite{AH}. The
two theories have the same Coulomb branch, but differ in the BPS spectrum as we
will discuss in section \ref{inst}, allowing us to distinguish them.

The third option (c) is shown after an $SL(2,\IZ )$ rotation. It has
parallel external legs, which are associated with 6d particles charged
under the global symmetry becoming massless when we take the bare
gauge coupling to infinity. It is not clear whether these particles
decouple from the low-energy 5d field theory or not. If they do, this
configuration describes a new 5d field theory which has the same
Coulomb branch as the pure $SU(2)$ gauge theory, but we will take the
conservative approach here and assume that such theories do not
exist. With this assumption we recover the two known possible pure
$SU(2)$ gauge theories. A similar question will arise below for
$SU(N_c)$ theories with $c_{cl} = \pm N_c$.

Note that generally, as described in section \ref{Webs}, local
deformations correspond to longitudinal deformations of the external
legs (namely, changing their length), so we may expect that
configurations where the deformations are excited will not be radiated
into the external legs (in the same way that a longitudinal wave is
not propagated on a string). However, there are presumably derivative
couplings of the 5d fields to the 6d fields, so the general question
of the decoupling of the 5d and 6d theories is still far from clear.

\begin{figure}
\centerline{\epsfxsize=130mm\epsfbox{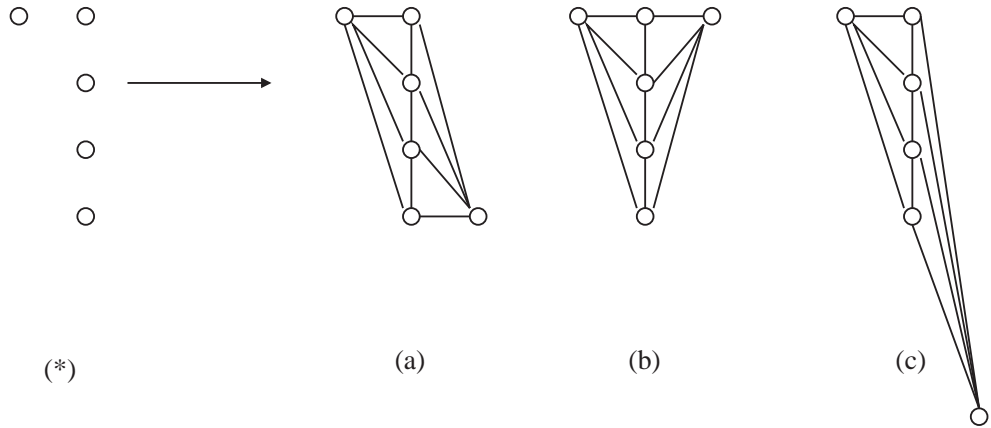}}
\medskip
\caption{The possible pure $SU(N_c)$ theories from Grids. Starting with a basic
Grid (*) we can construct $2N_c +1$ theories corresponding to the permitted
values for $c_{cl}$. $c_{cl}=0$ is shown in (a), the two Grids for
$\left| c_{cl} \right| = N_c$ are shown in (b,c).}
\label{sunc_options}
\end{figure}

Next, consider the possible configurations that result in \underline{a pure $SU(N_c)$ gauge theory}. Figure
\ref{sunc_options} shows the $N_c=3$ case. Similarly to the previous example we must have a vertical column of $N_c+1$
adjacent points, and two additional points, one on the left and one on the right. The left point can be fixed using the
residual $SL(2,\IZ )$ as shown in part (*) of the figure. We are left with $2N_c +1$ possibilities, that we identify
with the different possibilities for $c_{cl}$ (\ref{c-cl2}). Part (a) of the figure shows the $c_{cl}=0$ configuration,
while the two options for $\left| c_{cl} \right| = N_c$ are shown in parts (b,c). We can further use the $SL(2,\IZ )$
symmetry
together with a rotation
 to get a $\IZ_2$ symmetry of this spectrum corresponding to $c_{cl} \to
-c_{cl}$ (or charge conjugation).
Note that the two last configurations have parallel external legs and equality holds in the field theory constraint
(\ref{c-cl2}). It is not clear if such configurations correspond to five dimensional fixed points. This discussion can
be generalized to theories involving flavors.

\begin{figure}
\centerline{\epsfxsize=100mm\epsfbox{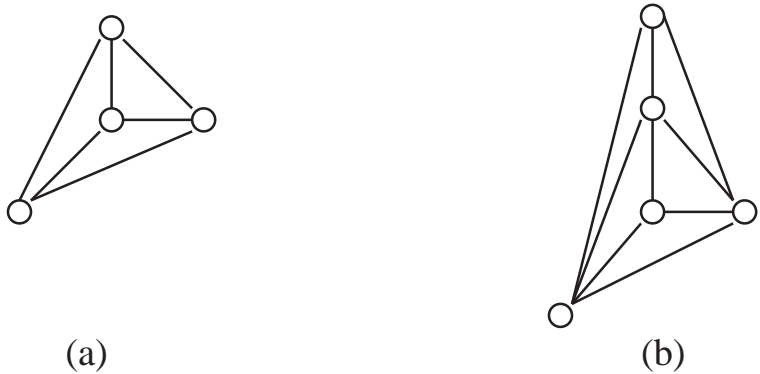}}
\medskip
\caption{Two Grids that do not have a SQCD type Lagrangian : (a) the
$E_0$ theory, (b) another example.}
\label{e0+}
\end{figure}

As a final example, consider two Grids that do not have a SQCD type
Lagrangian of $SU(N_c)$ with $N_f$ flavors, depicted in figure
\ref{e0+}. The first example (a) has a one dimensional Coulomb branch,
and is a deformation of the $E_0$ fixed point (figure \ref{e0}). The
second example (b) has a two dimensional Coulomb branch. In both
examples there are no global symmetries, and no parameters associated
with them, since there are only three external points. In section III
we will develop the tools to determine the BPS spectrum of these
configurations.

\subsection{The Higgs Branch}
\label{Higgs}

We call a configuration \underline{reducible} when it can be considered to consist of two independent Webs. When we
reach such a point in moduli space, we can separate the Web into sub-Webs, making use of the so-far ``unused'' 3
dimensions. In the process the rank of the local group is reduced and the $SO(3)_{``unused"} = Sp(1)_R$ symmetry is
broken. These are the roots for the Higgs branch\footnote{We thank
J. de Boer, K. Hori, S. Kachru, H. Ooguri, and Y. Oz
for discussions of this point.}. In field theory, we can go into the
Higgs branch either by turning on VEVs to fields which break the gauge
symmetry, or by turning on parameters (such as Fayet-Iliopoulos terms)
which force such VEVs to be turned on. We expect the former to
correspond to a local deformation of the brane configuration, but such
a deformation is not visible in our constructions, as
in any construction involving semi-infinite branes giving flavors
\cite{AH}. At the roots of the Higgs branch, we can however separate
the sub-Webs, and this corresponds to turning on some parameters which
force the low-energy theory into its Higgs branch.

\begin{figure}
\centerline{\epsfxsize=130mm\epsfbox{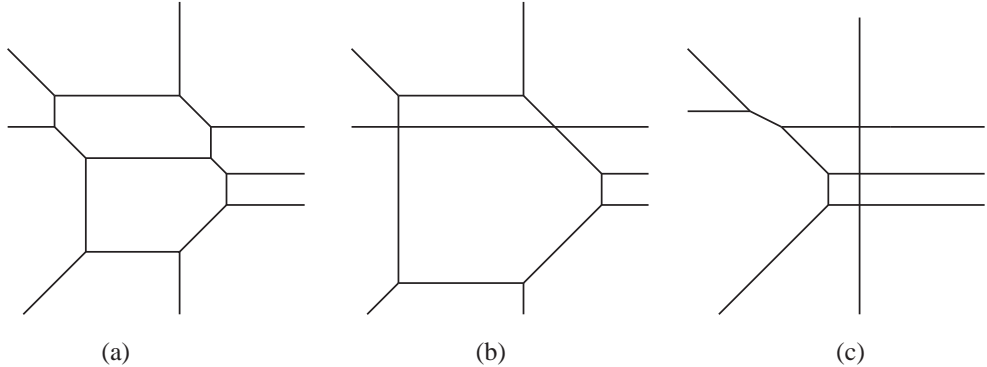}}
\medskip
\caption{Realizing the roots of the Higgs branch for a $N_c=3, N_f=4$ theory. (a) A generic
configuration, (b) a root for a mesonic branch, (c) a root for a baryonic
branch.}
\label{higgs}
\end{figure}

In SQCD theories, there are two general ways to enter the Higgs branch
: giving a VEV to a meson, and giving a VEV to a baryon. Consider a
configuration with $SU(N_c)$ gauge group, $n_L$ semi-infinite branes
(corresponding to quark flavors) on the left and $n_R$ semi-infinite
branes (also corresponding to quark flavors) on the right.
\underline{Giving a VEV to a meson} requires two semi-infinite branes,
one to the left and one to the right (figure \ref{higgs}(b)). These
semi-infinite branes have two quarks associated with them, from which
we can construct a gauge-invariant meson which can obtain a VEV. To
see this deformation, we need to first align the two ``flavor"
semi-infinite branes with one of the ``color" finite branes. Then, we
have a horizontal line that is an independent sub-Web. Having reduced
the Web, we find a root for a Higgs branch. In field theory terms this
process is described as follows : the masses of two of the quarks are
tuned (by setting their masses to be equal, and then moving on the
Coulomb branch) to be zero, and then we can turn on a parameter that
results in turning on a VEV for their meson operator. In the process
the rank of the gauge group is decreased by one.

In order to \underline{give a baryon a VEV}, we have to align $N_c$
semi-infinite branes from the same side (say right) with the $N_c$
color branes (figure \ref{higgs}(c)). This will separate the right
(vertical) support from the color branes. In this process we break the
whole gauge group. In the brane constructions of \cite{HW}, this
deformation was associated with a Fayet-Iliopoulos term, which
naturally results in a baryon obtaining a VEV. The interpretation in
our case is not clear, since the $U(1)$ part of the $U(N_c)$ gauge
group (associated with the Fayet-Iliopoulos term) does not seem to
exist in five dimensions.

We note that there is a difference between left and right
semi-infinite branes. It might be that the field theory knows to
distinguish between the respective two kinds of hypermultiplets. This
might be related to the fact that in 5d the mass of a hypermultiplet
is real, and it can have either a positive or a negative sign,
with a physical distinction between the two possibilities
\cite{W_phase}. Note that we can give either positive or negative
masses both to the hypermultiplets arising from left semi-infinite
branes and to those arising from right semi-infinite branes.

Let us determine the number of possible sub-Webs that can be separated from a given Web. Consider the list of 2d vectors
$(p,q)_i$ for the external legs. These vectors sum to zero (\ref{vcharge}). If any subset of these vectors sums to zero,
we can deform the Web so that it includes a separate sub-Web with these external legs, creating a disconnected
component. The lattice (partially ordered set) of zero-sum sub-Webs ordered by inclusion is a property of the asymptotic
configuration. Thus,
\be
\text{ $\sharp$(possible sub Webs) $= \sharp$(zero sum
subsets of the $(p,q)$ labels of external legs)}.
\ee
Once this condition for a sub-Web is satisfied, we still need to
tune global and local parameters to be able to turn on the deformation.

Other examples of moving apart sub-Webs also exist, which have no
obvious explanation in terms of mesonic or baryonic VEVs. For
instance, as noted in \cite{AH}, if we take the pure $SU(2)$ gauge
theory of figure \ref{su2_BPS} and take the gauge coupling to
infinity (and $\phi$ to zero), we get just a $(1,1)$ 5-brane
intersecting a $(1,-1)$ 5-brane, and we can separate the two. The
$SU(2)$ pure gauge theory is not expected to have a Higgs branch, but
at the strong coupling fixed point it is expected to have a Higgs
branch isomorphic to the moduli space of an $E_1 \simeq SU(2)$
instanton, which is $\IR^4/\IZ_2$ (generally, the strong coupling
fixed point of $SU(2)$ with $N_f$ flavors has an $E_{N_f+1}$ global
symmetry, and a Higgs branch equivalent to the moduli space of an
$E_{N_f+1}$ instanton). Thus, also in this case we can identify the
separation as corresponding to a Higgs branch, though it cannot be
expressed in terms of the variables of the $SU(2)$ gauge theory.

\newpage
\section{Instantons and Other BPS States in the 5d Theories}
\label{inst}

\begin{flushright}
{\it 'Tis true; there's magic in the Web of it.}\\
 Othello - Shakespeare.
\end{flushright}

\subsection{General Considerations and Some BPS States}
\label{gen_consid}

In this section we analyze the spectrum of BPS states of the five dimensional theories described in section
\ref{general}, and in their compactifications on a circle. Some of the BPS states of these theories were discussed in
\cite{AH}. There are several general types of BPS states in the 5d theories, corresponding to different types of central
charges as described below.
\begin{enumerate}
\item There are $W$ boson states in vector multiplets, which arise
from fundamental strings connecting the D5-branes as usual.
\item There are quark states or, in general, hypermultiplets, which
also arise from strings connecting D5-branes.
These strings will be found to differ from the Type IIB $(p,q)$ strings.
\item There are BPS saturated monopole strings, which arise from
D3-branes stretched along faces of the brane configuration.
\item As described above, in five dimensions instantons
can also be BPS saturated particles, charged under the global current
$J = *{\rm tr} (F \wedge F)$,
which reduces to the instanton number in four dimensions (we will call
these instantons also in five dimensions, hoping that this will not
cause too much confusion).
\end{enumerate}
When the gauge symmetry is non-Abelian, there exist finite size
instantons, and the instantons have a non--compact bosonic zero mode
corresponding to their scale size. This complicates the analysis of
the spectrum of instanton states (naively, it appears that there is a
continuum of particle states).  Therefore, we will discuss here only
the spectrum of instantons on the Coulomb branch, where the gauge
symmetry is Abelian. In this case there are no finite size instanton
solutions, and the instantons are singular gauge configurations. Since
they are localized, their properties depend on the short distance
physics \cite{Seiberg}, which in five dimensions is not determined by
the gauge theory. In our brane constructions, there is a well-defined
short distance theory (given by string theory), and we will be able to
compute the spectrum of instanton-like BPS states.

In 5d $N=1$ theories, like in 4d $N=2$ theories, BPS saturated states
can be either in vector multiplets or in hypermultiplets. The BPS mass
formula equates the mass of BPS saturated states with their central
charge. For particle states, the central charge is a combination of
contributions proportional to the charge of the state under global and
local $U(1)$ symmetries \cite{Seiberg}. In the pure $SU(2)$ gauge
theory it takes the form
\be
Z = n_e \phi + I / g_0^2,
\label{BPS}
\ee
where $n_e$ is the charge under the $U(1)$ gauge symmetry remaining on the Coulomb branch, $\phi$ is the scalar in the
$U(1)$ vector multiplet, $I$ is the charge under the global ``instanton number'' symmetry mentioned above, and $g_0$ is
the bare $SU(2)$ gauge coupling ($m_0=1/g_0^2$ may be viewed as a scalar in a background vector multiplet). The mass of
a BPS saturated state is proportional to $Z$. The standard gauge theory instanton has $I=1$. In theories with flavors,
there will be additional terms in (\ref{BPS}), corresponding to $U(1)$ subgroups of the flavor group. These will be
proportional to the quark masses, which again may be viewed as the scalars in background vector multiplets in the
adjoint of the global symmetry group.

In five dimensions the BPS charge is real, so a bound state of any two BPS
states with the same sign for the central
charge is a bound state at threshold. This makes the analysis of these bound states difficult, and we will not attempt
to completely determine their spectrum here, but just to discuss the states with the lowest possible charges (which
cannot be written as bound states of any other states). We call these states
\underline{simple BPS states}\footnote{The name comes from the analogy with the
simple roots of a Lie algebra, in the sense that they are indecomposable.}. The central
charge allows, in principle, for bound states at threshold to decay as we move in the moduli space. However, since we
are not able to determine the spectrum of bound states at threshold, we cannot say whether this actually occurs or not.

The BPS charge for strings is given by
\be
Z_m = n_m \phi_D
\ee
where $n_m$ is the magnetic charge (the charge under $*F$) and $\phi_D =
\del {\cal F} / \del \phi$. The tension of a BPS saturated string is
proportional to $Z_m$.

When we go down to four dimensions, there is an additional scalar in the vector multiplet associated with the Wilson
loop around the circle, and $\phi$ becomes complex. There is no longer a global ``instanton'' charge for particle
states, and the central charge is given by a combination of the complex scalar in the vector multiplet and its
electric-magnetic dual \cite{SW,SW2} (contributions from other global charges also still exist, and become complex when
we compactify).  Since the central charge is now complex, BPS states can be bound, and decay along marginal stability
curves where their binding energy goes to zero. In the M theory description, all the BPS states described above now
arise from membranes ending on the 5M-brane. States which arose from strings in 5d will now correspond (at least for a
large compactification radius) to membranes wrapped around the appropriate circle in M theory. The monopole string
states which arose from 3-branes will now become particles when wrapped around the circle of the Type IIB theory. The
3-branes wrapped around the Type IIB circle can be identified with unwrapped membranes in M theory.

\subsection{Instanton States and String Webs in the Brane Constructions}

The study of instanton states and their moduli spaces has been greatly
simplified since the realization \cite{Douglas} that a small instanton
inside a Dp-brane is the same as a D(p-4)-brane bound to the
Dp-brane. In the brane constructions considered here, we are looking
for small instantons in a theory where the gauge group comes from
D5-branes, so it is natural to look for instantons in configurations
of D-strings inside the D5-branes. The D5-branes live on a line
segment, and we would expect the D-strings stretching along the same
line segment to correspond to the ``instanton'' particles of the five
dimensional theory.

\begin{figure}
\centerline{
\epsfxsize=170mm
\epsfbox{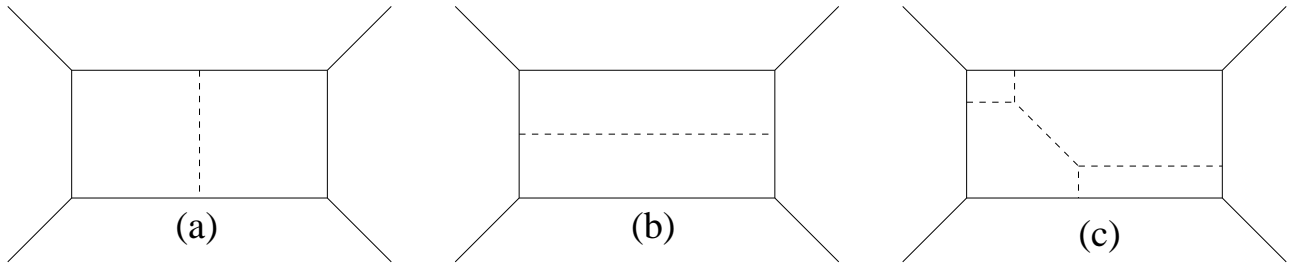}
}
\caption{A pure $SU(2)$ gauge theory. In this
figure and in similar figures in this section, 5-branes are denoted by
solid lines, and strings by dashed lines. Horizontal solid lines are
D5 branes and vertical solid lines are NS5 branes. In figure (a) there
is a W boson, given by a fundamental string stretched between two D5
branes. In figure (b) an instanton is given by a D-string stretched
between two NS5 branes. In figure (c) a bound state of a W boson and
an instanton is given by a Web of strings.}
\label{Xfig}
\end{figure}

The D-strings, being stretched along line segments, must end on some
other branes by charge conservation. However,
a D-string cannot end on a general $(p,q)$ 5-brane, but only on a NS 5-brane
(generally, a $(p,q)$ string, which may be
viewed as a bound state of $p$ fundamental strings and $q$ D-strings, can only end on a $(p,q)$ 5-brane \cite{ASY}).
Generically, our D5-branes will not end on NS 5-branes but on different branes, and even if they do end on NS 5-branes
these will only be NS 5-branes on one side of the D5-brane. Thus, we need to
look at more general configurations. The
simplest generalization of the usual stretched string states corresponds to a
Web of strings\footnote{It is interesting to note that similar configurations have
appeared in a recent paper \cite{GZ} in the context of enhanced $E_n$
symmetry in F theory.
However, we did not study the possible relation to the configurations in this work.},
analogous to the Webs of
5-branes discussed above, which live inside the internal faces of the brane configuration, and end on the 5-branes which
are its internal edges. Some examples of such configurations are drawn in figure \ref{Xfig}.

The principles for constructing such a Web of strings are completely analogous to the principles we used in section
\ref{general} when constructing a Web of 5-branes. A D-string inside the D5-brane breaks only half of the
supersymmetries of the brane configuration, so it is a BPS saturated state in the five dimensional theory. Similarly,
any $(p,q)$ string oriented in the $x,y$ plane such that
\be
\Delta x : \Delta y = -q : p
\ee
(this ratio is for a choice
of the Type IIB coupling $\tau=i$) will also break the same half of the supersymmetries of the brane configuration, so
we can construct BPS saturated states from any combination of such $(p,q)$ strings. As mentioned above, a $(p,q)$ string
can end on a $(p,q)$ 5-brane. With the orientations of the strings and 5-branes chosen as above to preserve
supersymmetry, a string will always be orthogonal to the 5-brane it ends on.
Physically, this assures \underline{zero force parallel to the 5-brane}, as required for static equilibrium.
In addition to ending on 5-branes, strings with charges $(p_i,q_i)$ can form vertices if $\sum p_i = \sum q_i = 0$
\cite{ASY}, just like 5-branes. As before, there will be no forces at such vertices (this is also a consequence of the
remaining supersymmetry). Any Web of strings with such vertices, and with $(p,q)$ strings ending on $(p,q)$ 5-branes,
will correspond to a BPS saturated state.

As an example of instanton configurations let us discuss the
brane configuration of figure \ref{Xfig}, described in the previous
section, which gives a pure $SU(2)$ gauge
theory at low energies. An obvious state here is the W boson,
corresponding to a fundamental string between the two D5-branes, as in
figure \ref{Xfig}(a). The mass of this state is equal (if we set the
string tension to one) to the separation $\phi$ between the
D5-branes. This is consistent, using the BPS formula (\ref{BPS}), with
the fact that it has $n_e=1$ and $I=0$.

However, there is a very similar state corresponding to a D-string between the
two NS 5-branes, as in figure \ref{Xfig}(b)\footnote{Note that this is the only
brane configuration corresponding to the pure $SU(2)$ gauge theory where this
D-string, corresponding to the naive gauge instanton as described above,
actually exists; we will discuss the general form of instanton states below.}.
The mass of this state is again the length of the corresponding string, which is
$1/g_0^2 + \phi$ (section \ref{Webs}). Thus, we find using equation (\ref{BPS})
that it has $I=1$ and $n_e=1$. It is clear that these instantons are in a vector
multiplet, since their configuration is related by S duality to that of the W
bosons\footnote{In the continuation of the theory past infinite coupling
discussed in \cite{AH} and in section \ref{Infinite}, the instantons and gauge
bosons interchange roles, and the instantons become the gauge bosons of the new
$SU(2)$ symmetry associated with the NS 5-branes.}.

In the brane configuration, a simple way to calculate the electric
charge $n_e$ of a state arising from strings is just to count the
number of boundaries that the strings have on the 5-branes, since
these correspond to electric charges inside the 5-branes. As described
above, the $U(1)$ gauge field of the low energy theory is really a
combination of the $U(1)$ gauge fields from all the different
5-branes, since the scalar component of that field was identified
with a shift in all the edges bounding a face in the brane
configuration in section \ref{Webs}. In our normalization,
every string boundary has a charge $1/2$.
Thus, a string Web configuration with $N_b$ boundaries has
\be
n_e = N_b/2.
\label{Icharge}
\ee
The instanton state (as well as the W boson configuration) has a
bosonic zero mode corresponding to its position inside the internal
face of the brane configuration. Since the face is compact, the
particle state corresponds just to the ground state of this bosonic
zero mode. This is the case in all the brane constructions of
particles discussed in this paper. The instantons (and the other
configurations) also have fermionic
zero modes, which turn them into multiplets of the five dimensional
$N=1$ supersymmetry.

Other configurations of string Webs breaking half of the remaining
supersymmetries also exist, which we would like to also interpret as
BPS states. An example of such a configuration is in figure
\ref{Xfig}(c).
This type of configuration can be deformed to become reducible as a collection of instanton and W states (one instanton
and one W in the figure). Since a bound state of these particles would be at threshold, it is hard to analyze whether it
exists or not.  The existence of string configurations where the instantons and W bosons are merged (as in figure
\ref{Xfig}(c)) may be an indication that such bound states indeed exist (for all $p,q \geq 1$). More complicated
examples of instanton configurations are presented below.

It should be emphasized that Webs of strings of this sort correspond to various different states in the 5d theories, and
not just to the states related to instantons of a non-Abelian gauge theory. In particular, states of this sort exist
even in theories like the $E_0$ theory discussed in \cite{Seiberg} (and constructed from 5-branes in \cite{AH}), which
have no non-Abelian gauge theory interpretation (i.e. no deformation leading from the fixed point to a non-Abelian gauge
theory with finite coupling; the low energy theory is still a $U(1)$ gauge theory).

\begin{figure}
\centerline{
\epsfxsize=60mm
\epsfbox{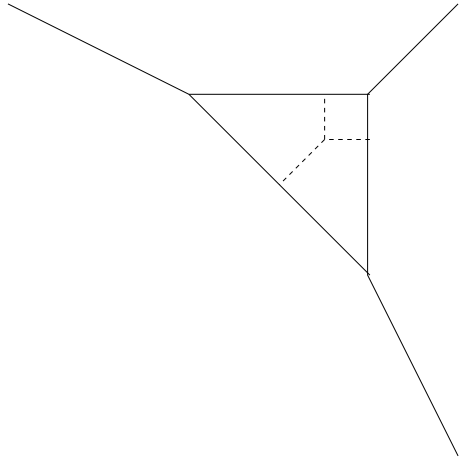}
}
\caption{A massive BPS vector multiplet on the Coulomb branch of the $E_0$
theory.}
\label{E0vect}
\end{figure}

The only BPS state corresponding to a string Web which seems to exist
in this theory is drawn in figure \ref{E0vect}, and it has $n_e=3/2$,
using the rule (\ref{Icharge}).  There is no global charge $I$ in this
case, and a direct computation confirms that the mass of this state is
$M = {3\over 2} \phi$. The fact that in this case there seems to be
no state with $n_e=1$ (which could be interpreted as a W boson) is
consistent with this theory not being related to any $SU(2)$ gauge
theory.

\subsection{BPS States for the Theories on a Circle}

In M theory descriptions of 4d $N=2$ gauge theories \cite{Witten}, the
brane configuration is described by a 5M-brane wrapped around some
Riemann surface, and the BPS states are given by membranes ending on
1-cycles inside the 5M-brane.  As discussed in
\cite{HennYi,Mikhailov}, these membranes are stretched along curves
which are holomorphic in a complex structure orthogonal to the one in
which the 5M-brane curves are holomorphic.  Physically, the condition
of an orthogonal complex structure assures that a membrane will end
orthogonally on the 5-brane, as expected of a minimal surface.  When
compactified on a circle, as described in section \ref{general}, our
theories are also described by wrapped 5M-branes, and the description
of BPS states will be similar. The only difference is that in our case
the surface on which the 5M-brane is stretched is embedded in
$\IR^2\times {\bf T^2}$ instead of $\IR^3 \times S^1$.  As conjectured
by Henningson and Yi \cite{HennYi}, and proven by Mikhailov
\cite{Mikhailov}, membranes with the topology of a disc will give rise
to hypermultiplet states while membranes with more boundaries can give
rise to either hypermultiplets or vector multiplets
\cite{Mikhailov}. It seems that states with the topology of a cylinder
always give rise to vector multiplets. Since the derivation of the
condition for a state to be BPS saturated is the same here as in
\cite{HennYi,Mikhailov}, we will not repeat it.

There is a simple relation between the local and global charges of a BPS state, appearing in the central charge formula,
and the topology of the 1-cycle which the membrane ends on. The 5M-brane is given by a two dimensional surface $S$
inside an ambient four dimensional space $M \simeq \IR^2 \times {\bf T^2}$. A BPS state is characterized by its boundary
1-cycle $c$. This has to be non-trivial in $H_1(S)$, otherwise it would shrink, but it must be trivial in $H_1(M)$ by
definition, since it is a boundary for the membrane.
These cycles are boundaries for the relative Homology $H_2(M/S)$ of membranes in $M$ with boundaries in $S$. So, we
require
\be
c\neq 0 \text{ in } H_1(S) \qquad c=0 \text{ in } H_1(M).
\ee
The genus of $S$ is
the number of internal faces in the Web, which is equal to the dimension $n_L$ of the Coulomb branch
(\ref{dim_coulomb}). We denote by $n_X$ the number of external legs,
and then $n_G=n_X-3$ is the number of global $U(1)$ charges.
The surface $S$ of genus $n_L$ has $n_X$ points removed (corresponding to the branes going out to
infinity). The Betti numbers are thus
\be
b_1(S)= 2n_L + n_X - 1 \qquad b_1(M) =2.
\label{bs_bm}
\ee
The natural injection $S \to M$, induced to homology, shows us that the
non-trivial 1-cycles in $H_1(S)$ that are trivial in $H_1(M)$ are a
subspace with dimension
\be
\text{dim(BPS state space)} = \text{dim}(H_2(M/S)) = b_1(S)-b_1(M)=2n_L+n_X-3
\ee
(we assume that not all branes are parallel, and thus the image in
$H_1(M)$ is indeed of dimension 2). This is exactly the dimension of
the space of possible charges upon reduction to four dimensions. There
are $n_L$ local gauge charges,
each one contributing both an electric and a magnetic charge appearing in the BPS mass formula (the magnetic charges
correspond to strings in the 5d theories). And, there are $n_G=n_X-3$ additional global $U(1)$ charges, using
(\ref{no_global}). Thus, we can identify each charge appearing in the BPS formula with a non-trivial 1-cycle in the
5M-brane, and the charge of a BPS state will be the number of times which the boundary of a membrane winds around this
particular cycle. Several examples of this will be given below.

For any combination of the cycles described above, we expect to have
BPS states in the theory corresponding to the minimal-area
configurations of membranes ending on that combination of
cycles. However, these states are not necessarily single particle
states, and may describe several particles. Only when we cannot
decompose a cycle $c$ into several cycles $c_i$ so that the mass of
the BPS state ending on $c$ is the sum of the masses of the BPS states
ending on the cycles $c_i$, we can be sure that there is indeed a
particle-like BPS state corresponding to this cycle.
We call such cycles \underline{simple cycles}, corresponding to the
simple BPS states defined in section \ref{gen_consid}.

We now turn to some examples. Our description of the 5M-brane configuration in section \ref{general} uses a complex
structure in which the holomorphic variables are
\be
s = \exp((x+ix_t)/L_t), t=\exp((y+iy_t)/L_t).
\ee
It will be convenient to describe some
of the BPS saturated membrane
configurations with the orthogonal complex structure given by
\be
{\tilde s} = \exp((x+iy_t)/L_t), {\tilde t}=\exp((y-ix_t)/L_t).
\ee
Our first examples are in the configuration of the pure $SU(2)$
gauge theory described in figure \ref{Xfig}, which,
upon compactification on a circle, is described by the curve
(\ref{su2_curve})
\be
As+t+ABst+Ast^2+ts^2=0.
\ee
First, let us describe the W boson of figure \ref{Xfig}(a). It is
given by a fundamental string stretched between two D5-branes, which,
upon compactification, becomes a membrane wrapped around one of the
cycles of the torus, which ends on two circles inside the 5M-brane. In
the 5d limit, we expect it to be given by a line with a constant value
of $x$. A possible configuration of this type is given by ${\tilde s}
= -1$.
Although two complex equations in 4 (real) dimensions intersect in a point in general, the complex equations for the
5-brane and the membrane (with complex structures orthogonal to each other) must be constructed to intersect along a
line.
It is easy to check that the intersection of this surface with the 5M-brane surface is indeed given, at least for large
$A$ and $B$ corresponding to a semi-classical region (or to the 5d limit), by two (topological) circles. They are given
by the solutions of
\be
\exp(y/L_t) = {1\over 2A} [AB + 2\cos(x_t/L_t) \pm
\sqrt{(AB+2\cos(x_t/L_t))^2-4A^2}].
\ee
The membrane wrapped around the component of the curve $\tilde{s}=-1$ which is between these two circles is the BPS
saturated state corresponding to the W boson. The sum of these two circles (with opposite orientations) is indeed
trivial in $H_1(M)$, and corresponds to one of the generators of $H_2(M/S)$, which corresponds to a state with $n_e=1$
and $I=0$. This state goes into the state of figure \ref{Xfig}(a) in the 5d limit. Similarly, a portion of the curve
$\tilde{t}=-1$ describes the instanton configuration of figure \ref{Xfig}(b).

As another example we can take the BPS saturated state of figure
\ref{E0vect}. The 5M-brane configuration, corresponding to the Grid
of figure \ref{more_grid}(b), may now be chosen to be
\be
1+st^2+s^2t-3Ast=0,
\ee
and the $n_e=3/2$ state is part of the membrane stretched along the curve $\tilde{s}+\tilde{t}=\tilde{s}\tilde{t}$. In
this case this membrane is the only non-trivial cycle in $H_2(M/S)$, so there are no states with a smaller electric
charge.

To describe other states, we will need to use different complex
structures, as in 4d $N=2$ theories \cite{HennYi,Mikhailov}. For
example, let us describe the BPS state corresponding to a 4d monopole,
which is a 5d monopole string wrapped around the compactification
circle. In the Type IIB string theory, this is a 3-brane stretched
along the rectangle of figure \ref{Xfig}. In M theory this is
identified with a
membrane stretched in a similar configuration. The equation for this
state is ${\rm Im}(s) = {\rm Im}(t) = 0$, which is holomorphic in a
complex structure corresponding to $s^\prime=x+iy, t^\prime=x_t-iy_t$.

\subsection{Hypermultiplet States and a First Apparent Paradox}
\label{Hypers}

\begin{figure}
\centerline{
\epsfxsize=170mm
\epsfbox{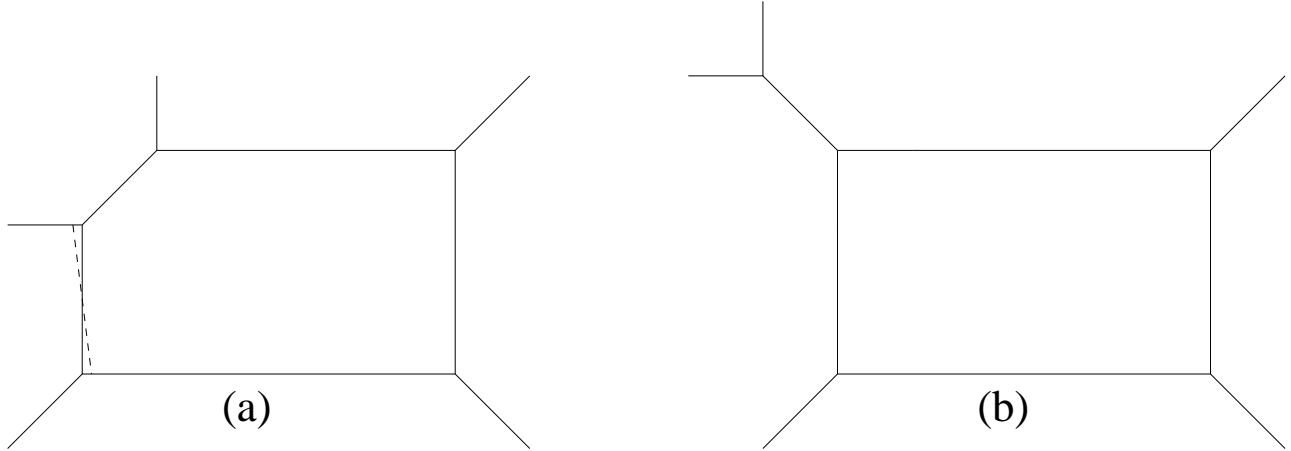}
}
\caption{The $SU(2)$ gauge theory with $N_f=1$. Figure (a) describes
the theory for low values of the mass ($|m| < \phi/2$), while figure (b)
describes the theory for large values of the
mass.}
\label{NfOne}
\end{figure}

In this section we will describe how to see some hypermultiplets in
the brane configurations, and this will lead us to an apparent
paradox, which will be resolved in the next subsection. The simplest
examples of hypermultiplets are quark states, which exist for instance
in the $SU(2)$ gauge theory with one quark ($N_f=1$). A brane
configuration for this theory was described in \cite{AH}, and it is
given in figure \ref{NfOne}.

In addition to the global $U(1)$ associated with the instanton number,
this theory has an additional $U(1)$ global
flavor symmetry. The BPS formula in this case is
\be
M = |Z| = |n_e \phi + I/g_0^2 + Q_f m|,
\ee
where $Q_f$ is the flavor charge and $m$ is the bare mass of the
quark. The standard gauge theory states in this case (which should
exist at least for weak coupling and low mass, i.e. in the
configuration described by figure \ref{NfOne}(a)) are the W bosons and
the two quark states. The W bosons have $n_e=1$ and $I=Q_f=0$ as
usual, so their mass is $M_W=|\phi|$, which is the distance between
the two D5-branes. This is a standard vector multiplet state, similar
to the ones
described in the previous subsection. The quark states, on the other
hand, are hypermultiplets which have $Q_f=1$,$I=0$ and $n_e = \pm 1/2$,
so their mass should be $M_Q = |m \pm \phi/2|$.

Before we identify these states, let us relate the parameters of the
brane configuration and the gauge theory in this case. The mass of the
W boson determines the distance between the D5-branes to be
$\phi$. The other parameters may be determined from the tension of the
monopole, which for $|m| < |\phi/2|$ is given by \cite{Seiberg,IMS}
\be
T_M = \phi/g_0^2 + 7 \phi^2/8 - m^2/2.
\ee
This follows from the general formula (\ref{calF}), which gives in this case
\be
{\cal F} = \phi^2 / 2g_0^2 + |\phi|^3 / 3 - |m+\phi/2|^3 / 6 -
|m-\phi/2|^3 / 6.
\ee
This determines the length of the bottom D5-brane in figure
\ref{NfOne} to be $1/g_0^2+\phi-m/2$, while that of the top D5-brane is
$1/g_0^2+\phi/2+m/2$.

It is natural to identify the quark states with strings going from the
``color'' D5-branes
to the ``flavor''
D5-brane. For one of the quark states, the corresponding string is drawn in figure \ref{NfOne}(a), and it has exactly
the correct mass given by the BPS formula, namely $M_Q = m+\phi/2$. When we compactify the theory on a circle, this
state becomes a membrane which is topologically a disc, as described in the next subsection. Thus, it corresponds to a
hypermultiplet \cite{Mikhailov}.

However, the identification of the second quark state is
confusing. Its mass is exactly the vertical distance between the top
D5-brane and the external D5-brane in figure \ref{NfOne}(a), but it is
not clear which string configuration gives rise to this state. A
$(1,-1)$ string stretched along the $(1,1)$ 5-brane seems to have
nothing to end on, and also would give a state of twice the desired
mass (a factor of $\sqrt{2}$ comes from the distance along the
diagonal, and another factor of $\sqrt{2}$ comes from the tension of
the $(1,-1)$ string). None of the configurations we have described so far
gives this state, so we have an apparent paradox. We will return to
this question, and resolve the paradox, in the next subsection.

\begin{figure}
\centerline{
\epsfxsize=170mm
\epsfbox{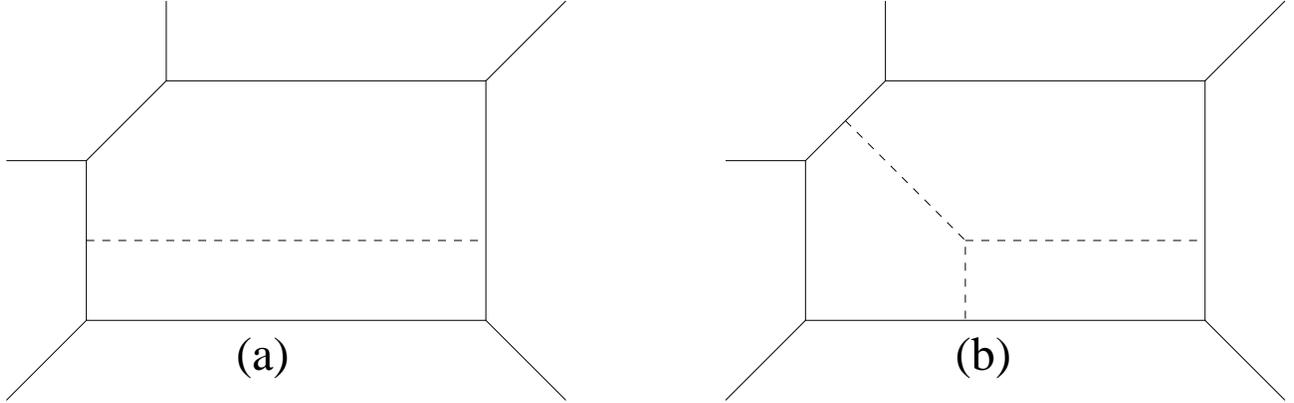}
}
\caption{Two instanton-like states in the $SU(2)$ gauge theory with $N_f=1$.
Figure (a) describes the $n_e=1$ state while figure (b) describes the
$n_e={3\over2}$ state.}
\label{NfOneInst}
\end{figure}

\subsubsection{Instantons and a Jump in the spectrum}
\label{jump}
Before resolving the paradox,
let us describe the instanton states in this theory. The simplest
state is given in figure \ref{NfOneInst}(a), and its mass is $M =
1/g_0^2+\phi-m/2$. From this we deduce that this state has $I=1$, as
expected, and $n_e=1$, which is consistent with the two boundaries of
the string in this case (\ref{Icharge}). The instanton state also has
a flavor charge $Q_f=-1/2$.  Presumably, this arises from quark zero
modes in the instanton background. Note that a state of this form
exists both for $m < \phi/2$ (figure \ref{NfOne}(a)) and for
$m > \phi/2$ (figure \ref{NfOne}(b)).

Another instanton state is drawn in figure \ref{NfOneInst}(b). As
drawn, this state exists only for $m < \phi/2$, and not for $m >
\phi/2$. How can we understand this jump in the spectrum ?
We will show that this BPS state disappears exactly when a quark becomes massless, allowing it to decay into other BPS
states, which do not disappear at this singularity in moduli space.
First, the quantum numbers of this state may easily be computed from the tensions and lengths of the corresponding
strings, and we find $I=1,n_e=3/2$ (as expected from equation (\ref{Icharge}) since the string has 3 boundaries) and
$Q_f=1/2$. These quantum numbers are the same as those of the instanton described in the previous paragraph
($I=1,n_e=1,Q_f=-1/2$), plus those of a W boson ($I=0,n_e=1,Q_f=0$), plus those of one of the quark states
($I=0,n_e=-1/2,Q_f=1$). For $m <
\phi/2$, the sum of the masses of these three states is larger than
the mass of the instanton in figure \ref{NfOneInst}(b), since the mass
of the quark is $M_Q = |m - \phi/2|$. However, for $m > \phi/2$, the
masses are the same, so this instanton can decay into these 3 states
(or to bound states of these, if they exist). The figures seem to
suggest that this decay indeed occurs, since we do not see this state
for $m > \phi/2$, but it is possible that the instanton state still
exists as a bound state at threshold even beyond this transition
point. Note that the transition point is exactly at $m = \phi/2$,
where one of the quarks becomes massless. However, there seems to be
no analog of the ``marginal stability curves'' of \cite{SW} in this
case.

As we change $\phi$ such that $\phi /2-m$ changes sign, a diagonal edge in the Web changes direction (figure
\ref{NfOne}). This is the edge that corresponds to the quark that has its mass change sign in this process. Viewing
this process in the Grid diagram we see that a diagonal line inside a square ``flops'' -- changes to the other diagonal.
This transition looks like a flop transition in Calabi-Yau spaces.

\subsection{Strips and Instantons in $(p,q)$ 5-branes}

Let us now return to the problem of the ``missing'' quark state. Naively, the appropriate state should be a $(1,-1)$
string inside the $(1,1)$ 5-brane. However, as discussed above, this has two problems. First, this string has nothing to
end on. Second, it has twice the required tension (for $\tau=i$). Since it does, however, seem natural that the state
should be stretched along the $(1,1)$ 5-brane, let us ask what string states exist inside the $(1,1)$ 5-brane. The
obvious stringy state there is the instanton of the $5+1$ dimensional gauge theory.
At first,
since the (small)
instanton in a D5-brane is identified with a D-string, we might think that the instanton in a $(p,q)$ 5-brane
is just a $(q,-p)$ string. However, in general this string does not have the correct tension to be the instanton of the
6d gauge theory, and for general values of $\tau$ it breaks all of the supersymmetry and not just half of it, so this
identification is not correct.

Let us first look at the gauge theory of a D5-brane. Some of the terms in the
low-energy theory of a D5-brane are (up to constants)
\be
{\cal L} \sim {T_s \over \lambda} F \wedge *F + \chi F \wedge F \wedge
F + B_{RR}
\wedge F \wedge F,
\label{daction}
\ee
where $F$ is the two-form field strength of the D5-brane, $B_{RR}$ is the RR 2-form, and $\lambda$ and $\chi$ were
defined in (\ref{tau}). From (\ref{daction}) we can read off the tension of the instanton, which is $T_s/\lambda$. For
$\chi=0$ the tension of the instanton is the same as the tension of a D-string, which also carries the correct charge
due to the last term in (\ref{daction}), so we can identify the two strings. In general, however, the tension of the
D-string is $T_{D1}=|\tau| T_s$, so it is not the same as the instanton tension (the general formula for the tension of
a $(p,q)$ string is $T_{p,q} = |p+q\tau| T_s$). To find the instanton tension for the $(1,1)$ 5-brane, we perform the
$SL(2,\IZ )$ transformations which act on $\tau=\chi/2\pi+i/\lambda$ as $\tau
\to -1/\tau$ and then $\tau \to \tau+1$.
We find that the gauge coupling of a $(1,1)$ 5-brane is given by
\be
{1\over g_{1,1}^2} = {T_s \over {\lambda \sqrt{(\chi/2\pi+1)^2
+ 1/\lambda^2}}},
\ee
while the tension of a $(1,-1)$ string is
\be
T_{1,-1} = T_s \sqrt{(1-\chi/2\pi)^2 + 1/\lambda^2},
\ee
so it is clear that the two objects are not the same.
For general $(p,q)$ 5-branes, and also for the
D5-branes at generic values of the axion, the instanton in the 5-brane
gauge theory is not related to any string in spacetime. In particular,
for $\tau=i$, the instanton in a $(1,1)$ 5-brane has half of the
tension of a $(1,-1)$ string, thus resolving one of the two problems
mentioned above.

In general, the tension of an instanton inside a $(p,q)$ 5-brane is
\be
T_{p,q}^{instanton} = {Im(\tau) \over \left|p+\tau q\right|}T_s.
\ee
So, for $\tau=i$, the tension of a $(q,-p)$ string is
\be
T_{p,q}^{string}=\left|q-\tau p\right| T_s = T_s\sqrt{p^2+q^2},
\ee
while the tension of an instanton inside a $(p,q)$ 5-brane is
\be
T_{p,q}^{instanton}={T_s \over \sqrt{p^2+q^2}}.
\ee

\begin{figure}
\centerline{
\epsfxsize=100mm
\epsfbox{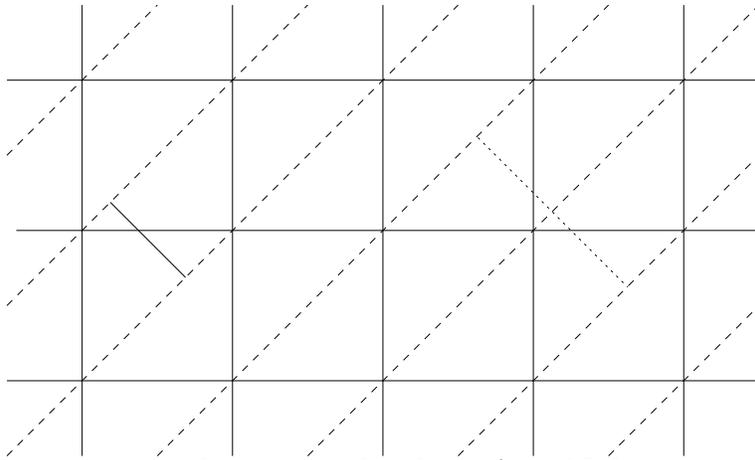}
}
\caption{5-branes, ``strips'' and strings in the plane of the M theory
torus. The dashed lines correspond to a $(1,1)$ 5-brane (drawn with
its images), the solid diagonal line corresponds to an instanton
``strip'' inside it and the dotted line corresponds to a $(1,-1)$ string.}
\label{oneone}
\end{figure}

It is easy to see all this in the M theory picture, which we get by
compactifying these theories on a circle as described in section
\ref{general}.  Drawing the torus for $\tau=i$ as a square Grid, a
$(1,1)$ 5-brane looks (in the plane of the M theory torus) like the
dashed diagonal lines in figure \ref{oneone}, while the $(1,-1)$ string
is a membrane stretched along the $(1,-1)$ cycle of the torus,
corresponding to the dotted line in figure \ref{oneone}. The
instanton, on the other hand, corresponds to a membrane stretched
along an interval between two adjacent images of the 5-brane, drawn as
the solid diagonal line in figure \ref{oneone}.  The other dimension of the
membrane is unwrapped, and it forms a string inside the 5-brane. Since
this membrane is stretched over an internal interval, we call it a
``strip''.

From figure \ref{oneone} it is clear that the instanton string has half the
tension of a $(1,-1)$ string, and that two instantons can form a $(1,-1)$ string
by joining together, and then they can leave the 5-brane. For a general value of
$\tau$ this process does not happen. Let us
\underline{determine the general condition for strips to join and
form a string}. First, let us assume that the strip is an instanton inside a
D5-brane. Its tension is $T_{1,0}^{instanton} = Im(\tau) T_s$. Then, it turns
out that such a process is possible if and only if $Re(\tau)$ is rational.
Denote
\be
Re(\tau)=a/b, \qquad a,b \in \IZ \text{  relatively prime}.
\ee
Then, $Re(-a+\tau b)=0$, and \underline{$b$ strips can join to form a $(-a,b)$
string}\footnote{For $b=1$ this process is familiar. For $a=0$ it is the usual
process of a small instanton leaving a D5-brane as a string, while for other
values of $a$ it is its image under the $\tau \to \tau+a$ transformation in
$SL(2,\IZ)$.} whose tension is $T_{-a,b}^{string}=bIm(\tau)T_s$. For a general
$(p,q)$ 5-brane we can perform an $SL(2,\IZ)$ rotation transforming the 5-brane
into a D5-brane, and use the previous analysis. Explicitly, denote by $(e,f)$ a
lattice vector that forms a basis together with $(p,q)$, that is $pf-eq=1$. In
this basis the new modular parameter is given by $\tau^\prime=(e+\tau f)/(p+
\tau q)$. The condition for the process to be possible is now that
$Re(\tau^\prime)=a/b$ is rational. If this is the case, $b$ strips inside the
$(p,q)$ 5-brane can join to create a $[b(e,f)-a(p,q)]$ string. This effect is
somewhat similar to what happens in the brane construction of 4d $N=1$ SYM
\cite{WittenNone}, where, in an $SU(n)$ gauge theory, $n$ MQCD strings can
combine into a Type IIA string, which can then leave to the bulk. However, since
in our case a rational condition is involved, the physical significance of this
process is unclear.

\begin{figure}
\centerline{
\epsfxsize=100mm
\epsfbox{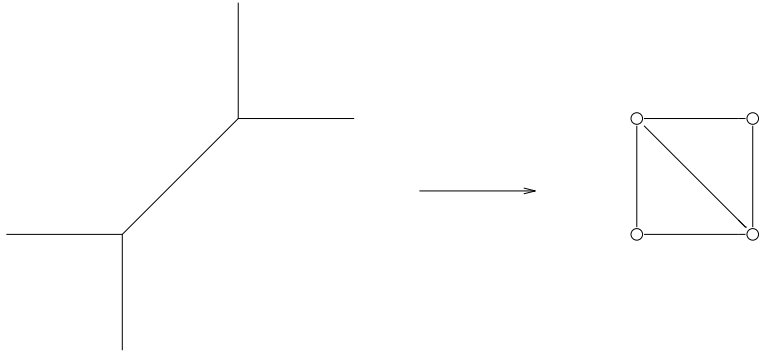}
}
\caption{The local region near a local $(1,1)$ 5-brane and its Grid
diagram.}
\label{local}
\end{figure}

Let us now go back to the problem of the missing quark. It is now
clear that an instanton stretched along the $(1,1)$ 5-brane in figure
\ref{NfOneInst} has the right mass to be the quark state, but we still
need to show that this instanton can end on the vertices which bound
this 5-brane. We will show that this is possible when the
configuration is wrapped on a circle, so that locally the $(1,1)$
5-brane and instanton look like figure \ref{oneone} and the 5-brane
configuration is described by a polynomial curve. Taking the 5d limit
we will find the configuration described above. We will give two
arguments for the existence of a configuration in which the instanton
ends on the vertices bounding the 5-brane -- a topological argument
and an explicit construction of the configuration. For both we will
analyze just the local region of the brane configuration corresponding
to the $(1,1)$ 5-brane and the branes it ends on -- clearly the
configuration is not expected to change significantly due to the
existence of other 5-branes far away. This local region and the
corresponding Grid diagram are drawn in figure \ref{local}.

First, a topological argument, which is a special case of the general topological arguments described in the previous
subsection. Topologically, the relevant region of the 5-brane surface in this case is a sphere with 4 holes
corresponding to the outgoing 5-branes, embedded in $\IR^2 \times {\bf T^2}$. There are, therefore, 3 topologically
non-trivial one-cycles in the 5-brane. Two (combinations) of these may be identified with the topologically non-trivial
cycles in the ${\bf T^2}$, leaving one topologically non-trivial cycle which a membrane can end on. By the arguments of
the previous subsection, there will be a BPS state ending on this cycle. In the particular case we are interested in,
for which the two pairs of branes emanating from the configuration are parallel, we can explicitly describe this BPS
state and see that it is indeed localized on the $(1,1)$ 5-brane. In this case the non-trivial 1-cycle is just a cycle
going around two of the parallel external branes, say the two D5-branes (a cycle going around the two NS 5-branes is
topologically equivalent to this). This is a finite circle going (more or less) around the $(1,1)$ 5-brane, and there
will be a membrane configuration ending on it. Moreover, since the boundary is topologically a circle, this membrane
will have the topology of a disc, so we are assured that this state will be a hypermultiplet as expected for the quark
state \cite{Mikhailov}.

Now, an explicit construction. The curve describing the 5-brane in
the configuration of figure \ref{local}, derived by the methods
described in section \ref{general}, is
$1+s+t+Ast=0$ (with $|A| < 1$; the length of the $(1,1)$ 5-brane in
the 5d limit is proportional to $-\log(|A|)$). A $(1,-1)$ string
parallel to the 5-brane is given by a membrane stretched on the curve
$\tilde{s} = \tilde{t}$. This intersects the 5M-brane along a closed
one-cycle which goes along the $(1,1)$ 5-brane segment, which is a
member of the non-trivial topological class discussed above. Part of
the membrane is enclosed by this cycle, and can be taken to be our BPS
configuration (it is BPS according to the general arguments described
above). It is easy to see that in the 5d limit, this state becomes
exactly a strip stretched along the finite $(1,1)$ 5-brane segment,
which has exactly the correct properties to be our ``missing'' BPS
state.

\subsection{BPS Spectra in Pure $SU(2)$ Gauge Theories and Another
Apparent Paradox}

In this subsection we will describe the most general brane
configuration corresponding to a pure $SU(2)$ gauge theory, and
analyze its BPS spectrum. The analysis of theories giving pure $SU(2)$
gauge theories was performed in section \ref{SUtwoGen}, resulting in
Grid diagrams having external vertices at the points
$(0,0),(1,m-1),(2,2m+n)$ and $(1,m+1)$ (up to global shifts), where
$m$ can be any integer, and $n=-1,0,1$ (as discussed in section
\ref{SUtwoGen}, the theories with $n=\pm 2$ do not seem to arise as
purely five dimensional theories, unlike the theories discussed here,
which may be reached by perturbing the 5d SCFTs corresponding to their
infinite coupling fixed point). There is an obvious $n \to -n$
symmetry, so we will only analyze here the cases of $n=0,1$. As argued
in section \ref{SUtwoGen}, the theories with different values of $m$
(and the same value of $n$) are expected to be the same at low
energies, since there is a residual $SL(2,\IZ )$ symmetry relating
them. However, we will see that the instanton states have a different
description in terms of strings for different values of $m$.

The theories with different values of $n$ behave rather differently, so we will
analyze each case separately. The $n=0$
series is characterized by having two parallel finite $(-m,1)$ branes, in
addition to the two parallel finite D-branes
($(1,0)$ branes). An example of this is the configuration of figure
\ref{Xfig}, corresponding to $m=0$. In all
these cases, a continuation of the theory past infinite coupling (as
discussed in section \ref{Infinite}) leads again to
an $SU(2)$ theory, arising from the $(-m,1)$ branes.

\begin{figure}
\centerline{
\epsfxsize=170mm
\epsfbox{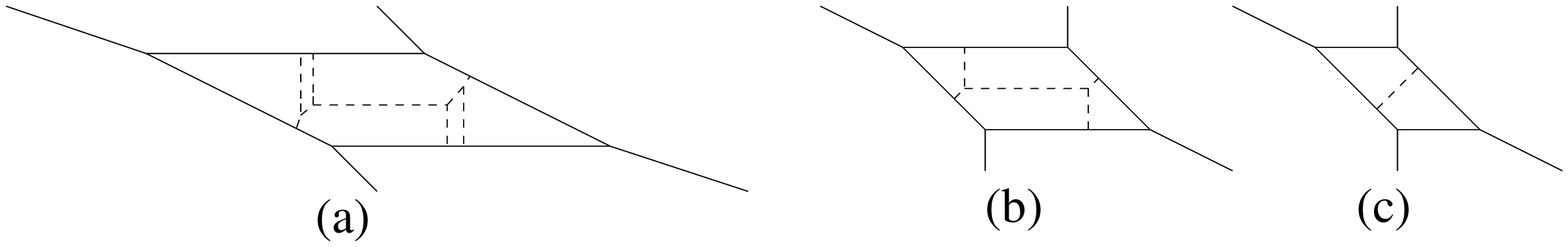}
}
\caption{Some string Webs corresponding to instanton states in $n=0$ models.}
\label{nzero}
\end{figure}

The spectrum of instantons that correspond to string Webs in these
theories depends on the relation between $1/g_0^2$ and $\phi$. For
weak coupling (namely, $1/g_0^2 \gg \phi$), the instanton
configuration with the
lowest electric charge in these theories which can be constructed as a
string Web is a generalization of the
instanton of figure \ref{Xfig}(b), depicted in figure \ref{nzero}(a)
for $m=-2$ and in figure \ref{nzero}(b) for $m=-1$. Configurations of
this type have $I=1$ and $n_e=|m|+1$. States with larger values of
$n_e$ may always be constructed by combining these states with W
bosons. For smaller $1/g_0^2$ (or larger values of $\phi$),
string Webs giving instanton configurations with smaller values of
$n_e$ also exist, such as the one depicted in figure \ref{nzero}(c)
for $m=-1$.

This last configuration has charges $I=n_e=1$ and a mass $M = 1/g_0^2
+ \phi$, but as drawn it exists only for $\phi > 1/g_0^2$. Similarly,
for any value of $m$, string Web configurations with $I=1$ and $n_e =
k < |m|+1$ exist only for $1/g_0^2 < \phi (|m| - 1 +
1/(|m|-k+1))$. One possible interpretation of this would be that at
this ratio of $1/g_0^2$ and $\phi$, these states decay into other BPS
states which they can be viewed as bound states at threshold of, as
described in section \ref{Hypers}. However, we argued above that the
theories for different values of $m$ should be equivalent, and in
any case no states of
appropriate charge seem to exist, even if we include ``strip'' states
of the form described above, so we run again into a paradox.

\begin{figure}
\centerline{
\epsfxsize=120mm
\epsfbox{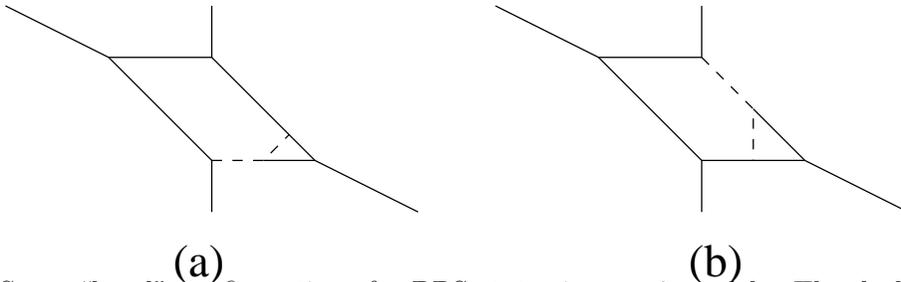}
}
\caption{Some ``bend'' configurations for BPS states in $n=0$
models. The dashed lines here correspond both to strings when they are
outside the 5-branes (a $(1,1)$ string in
figure (a) or a fundamental string in figure (b)), and to strings embedded
(as instantons) inside 5-branes (an instanton in the D5-brane in
figure (a) and in the $(1,1)$ 5-brane in figure (b)).}
\label{Bends}
\end{figure}

To resolve this paradox, we would like to argue that instead of
decaying, the state for $\phi < 1/g_0^2$ still exists, but it now
looks like
figure \ref{Bends}(a). We claim that such a configuration exists for
all $\phi$ and $1/g_0^2$, and is part of the moduli space of the
instanton configurations. When $\phi < 1/g_0^2$, ``bend''
configurations of this type are the only possible configurations for
the instanton state. The configuration of figure \ref{Bends}(a) looks
strange, since it has a string inside a 5-brane ending and turning
into a string outside the 5-brane. However, there is no contradiction
in this. The end-point of the string will look like a point-like
charge inside the 5-brane, and all the charges, when including both
``internal'' 5-brane contributions and bulk contributions, are conserved
in such configurations.

To prove the existence of a ``bend'' configuration of this sort we
will first show that locally a D-string inside a D5-brane can turn
into a $(1,1)$ string outside the 5-brane, as in figure
\ref{Bends}(a). A single D5-brane may be described by the equation
$t=1$ (in the usual coordinate defined above). We can now define a
string Web of a $(1,1)$ string and a $(-1,1)$ string going into two
overlapping $(0,1)$ strings by the equation
$\tilde{t}^2+2\tilde{t}+1-\tilde{s} \tilde{t}=0$. The D5-brane
intersects this string Web on a line, which separates it into two
regions, and each such region separately is a BPS separated
configuration which is exactly like the one we want. Namely, it
describes a single D-string (or instanton string) inside the D5-brane,
which smoothly leaves the D5-brane as a $(1,1)$ (or $(-1,1)$)
string. Using this sort of vertex, the rest of the configuration in
figure \ref{Bends}(a) involves things we have already seen, and we
claim that these configurations indeed exist and give rise to BPS
states (explicit configurations for these states may be constructed as
above).

A similar ``bend'' configuration, giving part of the moduli space of W-boson configurations, is described in figure
\ref{Bends}(b). Again, we can simply describe the non-trivial local vertex in this configuration, where an instanton
inside the $(1,1)$ 5-brane leaves it as a fundamental string. The $(1,1)$ 5-brane may be described by the equation
$s-t=0$. We can now put in a string Web corresponding to a fundamental string and a D-string turning into a $(1,1)$
string, described by the equation $\tilde{s}+\tilde{t}+1=0$. Again, the 5-brane separates the string Web into two parts,
and one of these looks exactly like the vertex of figure \ref{Bends}(b).
Note that bends satisfy the no-parallel-force condition.

We conclude that for all $n=0$ models (independently of $m$) there are
two basic BPS saturated states, the W boson with $n_e=1$ and $I=0$,
and the instanton with $n_e=1$ and $I=1$. All other states in these
models may be viewed as bound states of these two states, and we do
not know if bound states at threshold exist in this case or not.

\begin{figure}
\centerline{
\epsfxsize=100mm
\epsfbox{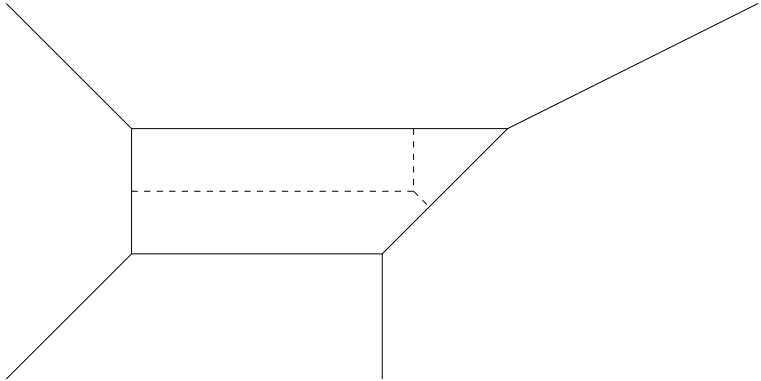}
}
\caption{A particular realization of $n=1$ theories, which are
deformations of the $\tilde{E_1}$ fixed point.}
\label{tildeEone}
\end{figure}

As discussed in section \ref{SUn}, the theories with $n=1$ are expected to be different, and to correspond to
deformations of the $\tilde{E_1}$ fixed point. We will verify here that these theories indeed have a different spectrum
of BPS saturated states, which we interpret as arising from a different discrete theta angle in $\pi_4(SU(2)) = \IZ_2$.
A particular realization of these theories is drawn in figure \ref{tildeEone}. Obviously, there is still the usual W
boson state with $n_e=1$ and $I=0$. Looking just at the bottom part of the diagram, we see that
it gives rise to a hypermultiplet since
(up to a 90 degree rotation) it is isomorphic to a configuration giving rise to a quark state (as described in section
\ref{Hypers}). In this case this state arises from a D-string inside the D5-brane. It is easy to check that the mass of
this state is $1/ g_0^2 + \phi/2$, so it has $n_e=1/2$ and $I=1$.
All other states, including the string Web state drawn in figure \ref{tildeEone}, may be viewed as bound states of
these two basic states. In this case we find that all states obey $n_e+{1\over 2}I \in \IZ $, in contrast to the $n=0$
case where all states obey $n_e \in \IZ $, so the spectra of the instanton states in the two theories are completely
different. Presumably, this quantization can be related to the different theta angle in the two theories.

\subsection{Comparison with Other Constructions}

The theories we analyzed in the previous section were constructed from
string theory (or M theory) in different ways in
\cite{Seiberg,MS,DKV,IMS}. We can compare the BPS spectrum that we
found above with the spectrum arising from these other constructions,
to verify that we are indeed constructing here the same low-energy
fixed points (despite the different high-energy regularization used in
the construction). Such computations were performed, for instance, in
\cite{MKV,KKV}.

The simplest comparison, which is the only one we will attempt here,
is with the construction of these theories as
low-energy limits of M theory compactified on Calabi-Yau manifolds,
where a Del-Pezzo submanifold shrinks to zero size. In this
construction, particle-like BPS states arise from membranes wrapped
around supersymmetric 2-cycles in the Del-Pezzo manifold, while BPS
saturated strings arise from 5M-branes wrapped around the whole
Del-Pezzo manifold.

For the $E_1$ theory and its deformations, the shrinking Del-Pezzo surface is isomorphic to a product of two ${\bf
CP}^1$'s. BPS states can arise from membranes wrapped around each of these ${\bf CP}^1$'s, which we identify with the W
boson and instanton state described above. Obviously, also in this construction both states are vector multiplets. Since
the BPS saturated string in this case arises from a 5-brane wrapped around both ${\bf CP}^1$'s, its tension should be
the product of the masses of the W boson and the instanton, and this is indeed the case for the states we found. Thus,
the BPS saturated spectra are the same in both cases (up to the possibility of having different marginally bound
states).

For the $\tilde{E_1}$ theory, the Del-Pezzo space is a ${\bf CP}^2$ blown up at one point. We recognize the blowup
parameter in our theory with the deformation parameter corresponding to the flow from the $\tilde{E_1}$ to the $E_0$
theory, described in \cite{AH} (and in section \ref{Infinite} below).  After this deformation, the W boson state
disappears, and only the instanton-like state of figure \ref{tildeEone} remains, which originally has $n_e=3/2$ and
$I=1$. Presumably, this state arises from a membrane wrapped around
the 2-cycle of
${\bf CP}^2$. The deformation to the $E_0$ theory occurs exactly when the length of the bottom D5-brane
in figure \ref{tildeEone} goes to zero, so we can recognize the hypermultiplet state described above (with $n_e=1/2$
and $I=1$) as arising from a membrane wrapped around the blown-up 2-cycle. The W boson state may then be identified as a
bound state of the $E_0$ state with a membrane wrapped around the blown-up 2-cycle in an opposite orientation. Note that
we expect membranes wrapped around blown-up 2-cycles to give rise to hypermultiplets, since the same blowing-up
procedure is used to add quark flavors in the geometrical construction. Thus, also in this case we find an agreement
between the different constructions.

\subsection{Enhanced Global Symmetries at the Strong Coupling Fixed Points}
\label{Enhanced}

The strong coupling fixed points of $SU(2)$ gauge theories with $N_f$
flavors are believed to have enhanced $E_{N_f+1}$ global symmetries at
their strong coupling fixed points (except for the $N_f=0$ case, where
two theories exist, one with an $E_1=SU(2)$ enhanced global symmetry,
and the $\tilde{E_1}$ theory without an enhanced global symmetry). In
our brane constructions of these theories, we cannot explicitly see
these enhanced global symmetries \cite{AH}. In fact, we cannot even
see the usual flavor symmetries when we have quarks coming from
semi-infinite D5-branes emanating in different directions. However,
when there is an enhanced global symmetry, the BPS states must fall into
multiplets of this symmetry, so we can check its existence using our
analysis of the BPS spectrum. We will do this here for some simple
examples.

First, let us take the $E_1$ theory. As discussed above, there are two
basic BPS states, which are both vector multiplets -- the W boson with
$n_e=1$ and $I=0$, and the instanton with $n_e=1$ and $I=1$. For
finite $1/g_0^2$, there is a $U(1)$ global symmetry associated with
the instanton number, and the masses of these two states are
different.  However, when $1/g_0^2=0$, these two states are
degenerate, and form a doublet of the enhanced $SU(2)$ at this point.
Note that this implies that the charge under the Cartan subgroup of
this $SU(2)$ is really $I-n_e/2$ and not $I$ as one might have naively
expected. At this value of $1/g_0^2$ and for any value of $\phi$, the
$SU(2)$ global symmetry is unbroken. All the states may be viewed as
bound states of the basic doublet states, and they fall into the
appropriate $SU(2)$ representations. When we turn on $1/g_0^2$, the
$SU(2)$ symmetry is broken to $U(1)$, and there is a mass splitting
proportional to $1/g_0^2$ between the states of each multiplet.

As another example, let us look at the $N_f=1$ theory depicted in
figure \ref{NfOne}. For arbitrary values of the couplings there is a
$U(1)\times U(1)$ global symmetry, which should be enhanced to $E_2 =
SU(2) \times U(1)$ at the strong coupling fixed point corresponding to
$1/g_0^2=m=0$. There are two basic vector multiplet states with
$n_e=1$ : the W boson with $n_e=1,I=0$ and $Q_f=0$ and the instanton
with $n_e=1,I=1$ and $Q_f=-{1\over 2}$ (using the conventions of
section \ref{Hypers}). We claim that these fall into a doublet of the
$SU(2)$ enhanced global symmetry, and are neutral under the remaining
$U(1)$. We can identify the Cartan generator of this $SU(2)$ with
$7I/8-n_e/2-Q_f/4$ and the $U(1)$ generator (up to normalization) with
$Q_f+I/2$. Some of the low-lying hypermultiplet states of these
theories were discussed in section \ref{Hypers}; we will analyze here
all the states which have $n_e=1/2$ (obviously, states in the same
multiplet of the global symmetry have the same value of $n_e$). There
is a quark state corresponding to an instanton in the NS 5-brane on
the left side, which has $n_e=1/2,I=0$ and $Q_f=1$. The other quark
state arises from a strip as described above, and has $n_e=1/2,I=0$
and $Q_f=-1$. There is also another hypermultiplet state similar to
the first quark state, arising from an instanton in the top D5-brane
-- this state has (by computing its mass) $n_e=1/2,I=1$ and
$Q_f=1/2$. The first and third state form a doublet of $SU(2)$ with
$U(1)$ charge $(+1)$, while the second state is a singlet with $U(1)$
charge $(-1)$. The $n_e=3/2$ state depicted in figure
\ref{NfOneInst}(b) is also a singlet of the $SU(2)$, with $U(1)$
charge $(+1)$.

\subsection{Singularities of the Curves}
\label{sing}

As in 4d $N=2$ theories \cite{SW,SW2}, we can associate singularities of the curves with massless charged BPS states. In
this section we will give some examples explaining how this works for our theories.

As the first example let us take the $E_0$ theory, whose brane
configuration and Grid diagram were given in figure \ref{more_grid}(b).
The curve read from the Grid diagram may be written as
\be
F(s,t)=1+st^2+s^2t-3Ast = 0.
\ee
The curve has a $\IZ_3$ symmetry acting as
$s\rightarrow \omega s, t\rightarrow \omega t, A\rightarrow \omega A$, with
$\omega$ a cubic root of unity. Presumably, this is related to the
$\IZ_3$ symmetry observed in the geometrical construction of this theory
in \cite{MS}. There are three singularities for this curve (solutions
of $F =\partial F / \partial s = \partial F / \partial t = 0$), which
are located at $A=\omega^i$, $i=0,1,2$.

In the 5d limit
$L_4 \to \infty$, and since $A \sim \exp(\phi L_4/2)$, we see that in
the $\phi$ plane all three singularities
coincide.
The BPS vector multiplet depicted in figure \ref{E0vect} has mass $3/2\phi$ while the BPS monopole tension is
$9/8\phi^2$. This is calculated from the simple geometry of the brane configuration. Both of these states are massless
if and only if $\phi=0$. When we compactify the theory, this singularity splits into 3 singularities (as noted in
\cite{GMS}), at each of which a different BPS saturated state is massless. After compactification, the central charge
for particles is given by a combination of the particle charge $n_e$ and the magnetic charge $n_m$ (associated with
monopoles wrapped around the circle). At each of these singularities, a state with some values of $n_e$ and $n_m$, which
may be viewed as a bound state of the two 5d states, becomes massless.

Next, let us look at the pure $SU(2)$ gauge theory with the curve
given by (\ref{su2_curve}). The singularities of this curve occur when
$2A \pm 2 \pm AB = 0$. According to (\ref{relation}), for finite
$1/g_0^2$, $A \to \infty$ in the 5d limit, and we find singularities
at $B = \pm 2$, which we identify with the point $\phi=0$ where the W
boson and the monopole (as well as the instanton) are massless. For
generic compactification radii, there are 4 separate singularities. In
the 4d limit, two of these singularities go off to infinity, and the
other two become the massless monopole and massless dyon singularities
of the 4d $N=2$ pure $SU(2)$ gauge theory \cite{SW}. In other examples
we can similarly reproduce the known singularity structure (described
in \cite{GMS}).


\section{Flows in Parameter Space Beyond Infinite Coupling}
\label{Infinite}

In this section we use the methods developed in section \ref{general} to study
some flows in parameter space of the five dimensional theories. As described in
section \ref{general}, a brane configuration describing a SQCD theory with $N_f$
flavors has $N_f+1$ real parameters (associated with the global symmetry whose
rank is $N_f+1$). These may be identified with the gauge coupling parameter
$m_0=1/g_0^2$ and the $N_f$ masses of the quarks. In gauge theories, obviously
$m_0$ is always positive. However, in the brane configurations all the
parameters are just transverse positions of branes, which may take any (real)
value. Thus, we can ask what happens when we deform the brane configuration so
that $m_0$ becomes negative. More generally, in a brane configuration
corresponding to a product of several groups, each one will have a gauge
coupling parameter associated with it, and the region of parameter space where
these are all positive is only a subset of the parameter space. These
``continuations past infinite coupling'' in five dimensions were first described
in \cite{AH}, and we elaborate on them in this section using the methods
developed above. In three dimensions such continuations were first described in
\cite{HW}, in four dimensions they were used to describe Seiberg duality in
\cite{EKG} and in two dimensions to describe level-rank duality in \cite{HH}.

Generally, when we take a parameter corresponding to a gauge coupling
to zero (taking the gauge coupling to infinity), the theory no longer
has an interpretation as a gauge theory. If the other parameters and
moduli of the theory are also set to zero, it will usually be at a
fixed point of the renormalization group flow, corresponding to some
superconformal field theory. There are two possibilities for the
behavior of the theory after we continue to flow in parameter space so
that $1/g_0^2$ becomes negative, and we will give several examples of
both. One possibility, described in the next subsection, is that the
theory after this deformation has an interpretation in terms of a
different gauge theory with a positive value of $1/g_0^2$. This means
that starting from the fixed point at $1/g_0^2=0$, we can flow to
either of the two gauge theories. However, we can also interpolate
between the two theories when we are at some point on the Coulomb
branch, and then the transition between them is generally smooth, and
does not encounter any singularities. This enables us to relate
various properties of the two theories, as described in the next
subsection. The other possibility is that the theory after the
continuation has no gauge theory interpretation, for instance it can
be the $E_0$ theory described above.
This possibility will be described in subsection \ref{others}.

\subsection{Continuation Past Infinite Coupling to Another Gauge Theory}

In this section we discuss continuations past
infinite coupling which lead to gauge theories. We will give several
examples of this phenomenon, and describe the relation between the
parameters, Higgs moduli and BPS states in the pairs of gauge theories
which are related in this way.

In the previous sections, we analyzed non-Abelian five dimensional
gauge theories in which the non-Abelian gauge symmetry comes from
several parallel D5-branes. Generally, however, parallel 5-branes of
any kind give rise to a non-Abelian gauge symmetry (the configurations
are all related by the $SL(2,\IZ)$ U-duality group). Recall that on
the Coulomb branch the Abelian gauge symmetry
actually comes
from a combination of the Abelian gauge fields on various different
5-branes, and not just from the D5-branes.

When we got a non-Abelian
gauge theory in the previous sections, the parameter $1/g_0^2$ was
proportional to the length of the parallel D5-branes at the point in
moduli space where they all overlapped. When we go to infinite
coupling, the length of the D5-brane segment goes to zero size. If we
continue changing the parameters (moving the asymptotic 5-branes), the
5-branes will rearrange themselves in some (generally) different
configuration, which in general will no longer have parallel
D5-branes. However, in some cases it will still have parallel branes
of a different kind, and in these cases we can find a different gauge
theory interpretation for the theory after the continuation past
infinite coupling.

\begin{figure}
\centerline{\epsfxsize=130mm \epsfbox{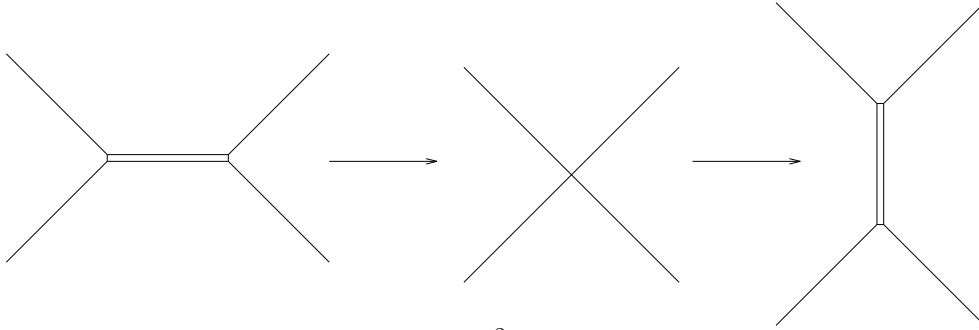}}
\caption{Changing the parameter $m_0=1/g_0^2$ in a pure $SU(2)$ gauge
theory, from a positive value to zero to a negative value.}
\label{su2_continuation}
\end{figure}

The simplest example of this phenomenon (mentioned already in \cite{AH}) arises in the pure $SU(2)$ gauge theory with no
theta angle (related to the $E_1$ fixed point). The simplest brane configuration related to this theory was described in
figure \ref{su2_BPS}. The theory has two basic BPS states, the W boson with a mass $m_W=\phi$ and the instanton with a
mass $m_I=1/g_0^2+\phi$. The only parameter of this theory is $m_0=1/g_0^2$, which may be changed by moving any one of
the asymptotic 5-branes. As described in figure \ref{su2_continuation}, when we take this parameter to zero we get the
$E_1$ fixed point, while if we take this parameter to be negative, we find two parallel NS 5-branes at the origin of the
new Coulomb branch. We can interpret these parallel 5-branes as giving rise to a new $SU(2)$ gauge theory, in which
the roles of the W boson and the instanton are interchanged. Thus,
equating the masses in the original and in the new theory by
$m_{\tilde{W}} = m_I$ and $m_{\tilde{I}} = m_W$,
we find that the parameters of the two theories are related by $1/{\tilde g}_0^2 =
-1/g_0^2$ and $\tilde{\phi} = 1/g_0^2 + \phi$ (this can also easily be
seen geometrically). Thus, away from the fixed point there is only one
gauge theory interpretation with a positive $1/g_0^2$, and the other
interpretation has no meaning as a gauge theory. Starting from the fixed point, we can
deform in two different ways, which in this case both lead to a pure
$SU(2)$ gauge theory
(this is related to the fact that
there happens to be an enhanced global symmetry at this point),
but generally they will lead to different theories.

If we connect the theories at $\phi=0$ (as drawn in figure
\ref{su2_continuation}) we pass through a singularity at the fixed
point, but we can also connect the two theories at a finite value of
$\phi$, and then the passage between them is completely smooth. Since
the passage is smooth we expect to have the same BPS spectrum in the
brane construction on both sides of the transition. Thus, we find that
the BPS spectrum of the $SU(2)$ gauge theory is invariant under the
exchange of W bosons and instantons. In fact, for this particular case
this follows from the enhanced global symmetry of the $E_1$ fixed
point, as described in section \ref{Enhanced}. However, in general we
will be able to relate in this way the BPS spectra of different
theories which have no apriori relation.
Note that if there are transitions in the BPS spectrum when we change the parameters, as suggested in section
\ref{jump}, we may not be able to relate in this way the spectra of the two theories at weak coupling, but only at very
strong coupling, when $1/g_0^2$ is much smaller than any other mass scale (such as $\phi$).

\begin{figure}
\centerline{\epsfxsize=70mm \epsfbox{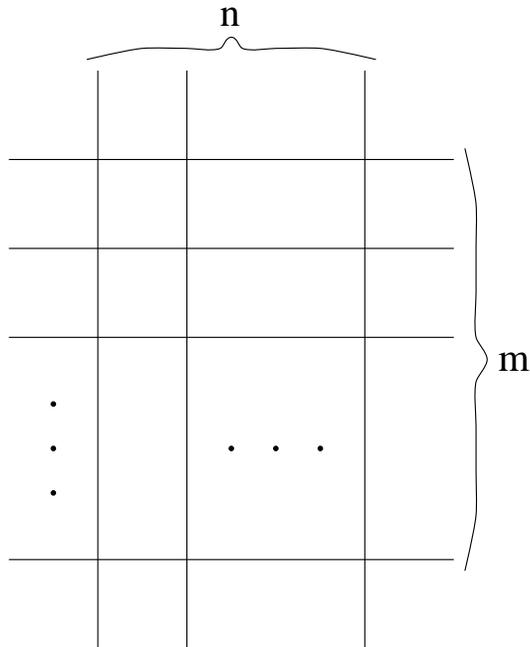}}
\caption{A Web of
5-branes which forms a product gauge group.  There are $n$ vertical lines
representing a NS5 brane and $m$ horizontal lines representing D5 branes.}
\label{rug}
\end{figure}

There are many other cases in SQCD theories where a continuation past
infinite coupling leads to a different gauge theory. A
particularly simple case is described in
figure \ref{rug}.  This configuration is slightly problematic, since (as in $SU(N_c)$ theories with $N_f=2N_c$ which
are a special case of these configurations with $n=2$) it is not clear if it has a strong coupling fixed point which can
be used to define it as a five dimensional field theory or not. In the brane configuration this issue is related to the
appearance of parallel external legs. However, we will ignore this potential problem here and assume that the theory
exists as a five dimensional theory, since the qualitative results will hold also for any other field theory, where the
description of the continuation past infinite coupling is slightly more complicated (but presents no fundamental
difficulties)\footnote{From the Grid diagram point of view it is clear that any Grid diagram may be reached by flowing
from a strictly convex diagram, taking the coefficients of some of the points to zero, so we can always find some
definition for any Grid diagram / brane configuration in terms
of a limit of a
five dimensional theory.}.

Interpreting the gauge group in figure \ref{rug} as arising from the horizontal D5-branes is natural when the
horizontal segments are much longer than the vertical segments, so that all the gauge coupling parameters associated
with the D5-branes are positive. In this case the gauge group is $SU(m)^{n-1}$. There are $n-2$ hypermultiplets in an
$\bf(m,m)$ bifundamental representation of each pair of adjacent gauge groups. In addition, there are $m$ fundamental
hypermultiplets for the first gauge group and $m$ fundamental hypermultiplets for the last gauge group, so that each
$SU(m)$ group has a total of $2m$ fundamental hypermultiplets.

In this case there are $n-1$ gauge coupling parameters, and one can
look at theories where some of them are negative and some of them are
positive, but we will discuss here only the deformation which takes
all of them to be negative, which arises by shrinking the diagram in
the horizontal direction and stretching it in the vertical direction
(so that the vertical segments are much longer than the horizontal
segments). In this configuration it is natural to associate the
non-Abelian gauge factors with the NS 5-branes, since these will have
positive gauge coupling parameters. In fact, the configuration we find
is exactly a 90 degree rotation of the configuration we started with,
if we exchange $n$ and $m$. Such a rotation does not change the
low-energy physics. Thus, the theory after we continued past infinite
coupling has an interpretation as an $SU(n)^{m-1}$ gauge theory, with
$m-2$ bifundamentals in $\bf(n,n)$ representations, and $n$ more
fundamental hypermultiplets for the first and last $SU(n)$ factors.
There are also other regions of the parameter space where neither of
these two interpretations is appropriate, but we will not discuss them
here.

Next, let us see how the various parameters and moduli of the two
theories are mapped to each other. In both cases the gauge symmetry at
a generic point on the Coulomb branch is $U(1)^{(m-1)(n-1)}$, and
there are $(m-1)(n-1)$ scalars labeling the position on the Coulomb
branch. The relation between the parameters of the two theories is
more interesting. The total number of external 5-branes is $2(n+m)$,
so there are $n_G=2(n+m)-3$ real parameters for both theories. In the
original interpretation, there are $n-1$ distances $L_i$ between
adjacent external NS5-branes, related to $n-1$ gauge coupling
parameters for the $SU(m)$ gauge groups. There are $2m$ vertical
positions of the external D5-branes, associated with the bare masses
of the $m$ fundamental hypermultiplets of the first and last $SU(m)$
groups. The remaining $n-2$ parameters correspond to bare masses for
the bifundamental hypermultiplets.

The exact correspondence between
the brane positions and the mass and coupling parameters is more
complicated, and we will not go into it here. However, it is clear
that in the continuation past infinite couplings, parameters which
were originally associated with gauge couplings become associated with
masses and vice versa. In general each parameter of the original
theory, a gauge coupling or a mass, will be some linear combination of
the gauge coupling parameters and masses of the other gauge theory. An
important property of this transformation is that, as above, there is
never a case in which a single configuration has an interpretation as
two different theories with positive gauge coupling
parameters. Starting from a point where all the branes intersect at a
single point, corresponding to the infinite-coupling zero-mass limit
of both theories, we can deform the theory into either one of the two
gauge theories.  Note that at this point the global symmetry is
manifestly
enhanced to $SU(n)\times SU(n)\times SU(m)\times SU(m)\times
U(1)$. For small values of $n$ or $m$ the enhanced global symmetry
will be larger (for $n=2$ it will include an $SU(2m)$ factor).
The BPS spectra of the two theories will again be related, as
described above.

We can also examine the deformations of these theories which move them
into their Higgs branches, as described in section \ref{Higgs}.
Figure \ref{rug} describes a point on the Coulomb branch of the
theories from which various Higgs branches emanate, which have
different interpretations in the two gauge theories related by the
continuation past infinite coupling.
One possible Higgs branch deformation involves the connection of $n-1$
adjacent segments of D5-branes with two semi-infinite D5-branes (one
from each side) into a single infinite D5-brane. After this
connection, this D5-brane forms a sub-Web which
can separate from the system,
 moving the low-energy theory into the
Higgs branch. In the original gauge theory interpretation in terms of D5-branes, this looks like a mesonic deformation
from the point of view of each of the $SU(m)$ gauge theories (as described in section \ref{Higgs}). Thus, we associate
it with a deformation which results in an expectation value for a gauge-invariant field of the form $Q^{1} B_1^2 B_2^3
\cdots B_{n-2}^{n-1} Q_{n-1}$ where $Q^1$ and $Q_{n-1}$ are fundamental hypermultiplets charged under the first and last
$SU(m)$ factors, and $B_i^j$ is a bifundamental charged under the $i$'th and $j$'th $SU(m)$ groups. Such a deformation
breaks each of the $SU(m)$ gauge groups to
$SU(m-1)$, as is obvious from the figure.

The interpretation of this deformation from the point of view of the
other gauge theory interpretation is completely different. If we
remove the $i$'th D5-brane (we assume $1<i<m$), this now affects only
the $i-1$'th and $i$'th $SU(n)$ groups, and we may associate this
deformation with the $i-1$'th bifundamental field getting an
expectation value proportional to the unit matrix, which breaks
$SU(n)_{i-1} \times SU(n)_i$ to a diagonal $SU(n)$. The gauge
invariant operator getting a VEV in this case is $(B_{i-1}^i)^2$.
If we move a D5-brane from one of the two ends ($i=1$ or $i=m$), the interpretation as a bifundamental getting an
expectation value is replaced by a baryon constructed from the $n$ ``external'' quarks getting an expectation value,
exactly as described in section \ref{Higgs}. The deformation of separating an NS 5-brane has the opposite interpretation
in each of the two theories.

Thus, the continuation past infinite coupling may be used also to
relate the Higgs branch of theories, which apriori have no
relation. However, since we cannot see the whole Higgs branch in the
brane configurations, this does not imply an exact equality between
the Higgs branches of theories related in this way. In particular, the
dimensions of the total Higgs branches of the two gauge theories
described above are not the same if $n \neq m$. This is analogous to
the continuation past infinite coupling in four dimensional theories
(described, for example, in \cite{EKG,EGKRS}), where the continuation
is smooth if it is performed on the Higgs branch. There, the Higgs
branches match between the two theories, while the Coulomb branches do
not generally match when the continuation is performed for 4d $N=2$
theories.

\subsection{Flows to Theories with no Gauge Theory Interpretation}
\label{others}

In this subsection we describe some deformations which lead to theories
with no gauge theory interpretation. The prototype of such transitions
is the transition found in \cite{Seiberg} and described in the brane
configurations in \cite{AH} which connects the $\tilde{E_1}$ theory
with the $E_0$ theory. We will focus here on generalizations of this
example to $SU(N_c)$ gauge theories, though many other examples also exist.

\begin{figure}
\centerline{ \epsfxsize=130mm \epsfbox{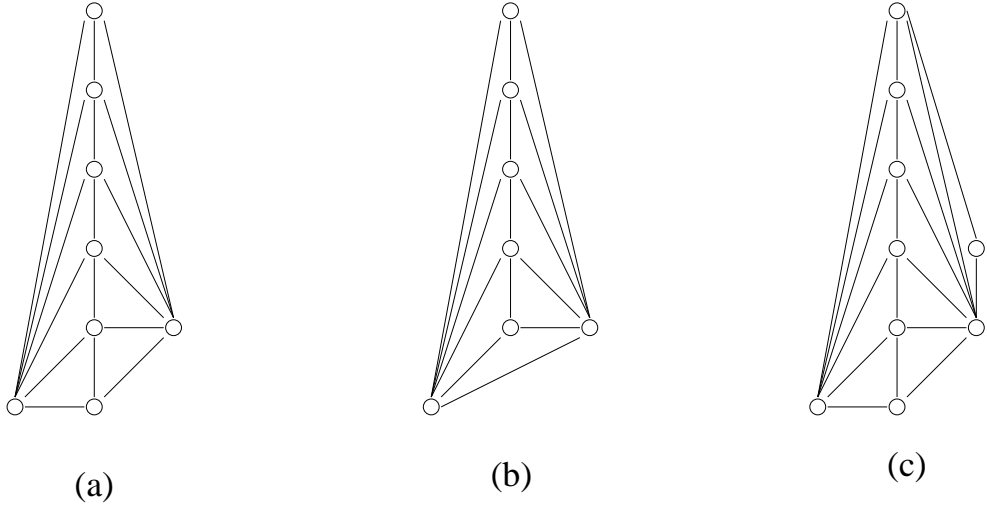} }
\caption{$E_0$-like transitions.  These are Grid diagrams which
represent various stages in a transition to $E_0$-like theories.
Figure (a) describes the Grid diagram for a pure $SU(N_c)$ theory with
$c_{cl}=N_c-1$ (in the case of $N_c=5$). In figure
(b) we flow to an $E_0$-like theory. Figure (c) describes adding a
flavor to the theory.}
\label{new_ezero}
\end{figure}

We can start describing this deformation in the
configuration
whose Grid diagram is given in figure \ref{new_ezero}a. For $N_c > 2$,
such a diagram describes the $SU(N_c)$ theory with $c_{cl}=N_c-1$, as
described in section \ref{SUn}. For $N_c=2$
the same figure describes the theory with a non-zero theta angle
related to the $\tilde{E_1}$ fixed point.
The curve which describes the
theory is read off from the Grid diagram along the lines described in section
\ref{general}.  It is
\bea
1+s^2t+s(t^{N_c}+b_{N_c-1}t^{N_c-1}+\cdots+b_1t+b_0) = 0.
\label{pure}
\eea
$b_0$ is related to the gauge coupling and is
the only parameter of the theory, while the other parameters of the
curve are related to the $N_c-1$
moduli of the Coulomb branch.

Taking the gauge coupling parameter to large negative values
corresponds to taking $b_0 \to 0$, and the resulting theory is drawn
in figure \ref{new_ezero}b. Now there are only 3 external legs, so
there are no longer any parameters for the theory, and it has no gauge
theory interpretation, like the $E_0$ theory.

In a similar way we can use the curves to describe any other flow in
the parameter space of these theories. For instance, we can add a
quark flavor to the theory in figure \ref{new_ezero}(a), resulting
in the theory of figure \ref{new_ezero}(c). The curve describing
this theory is now
\be
1+s^2t+cs^2t^2+s(t^{N_c}+b_{N_c-1}t^{N_c-1}+\cdots+b_1t+b_0) = 0.
\label{onef}
\ee
The constant $c$ is related to the mass of the hypermultiplet which
was added.  When $c$ is very small the curve (\ref{onef}) reduces to
the curve (\ref{pure}).  This corresponds to taking the mass of the
hypermultiplet to positive infinity.
We can deform to another flavor-less theory by taking the mass
parameter to negative infinity and ``integrating out'' the quark
flavor. In the curve, this is done by taking $c\to \infty$, and
rescaling the variables of (\ref{onef}) so that the equation remains
finite. In this case the required scaling is ${\tilde t} =
c^{-1/(2N_c-2)} t$, ${\tilde s} = c^{N_c/(2N_c-2)} s$ and $\tilde{b_i}
= c^{(i-N_c)/(2N_c-2)} b_i$. In terms of the new variables the curve
is now
\be
1+\tis^2 \tit^2 + \tis(\tit^{N_c} + \tib_{N_c-1} \tit^{N_c-1} + \cdots +
\tib_0) = 0.
\ee
This is exactly the curve of the theory with $c_{cl} = N_c-2$. Generally, by adding a quark with positive infinite mass
and taking its mass to negative infinity we can reduce the value of $c_{cl}$, while an opposite flow increases the value
of $c_{cl}$ (these phase transitions were described in the geometrical construction of these theories in
\cite{W_phase}). Of course, we are always limited to $|c_{cl}| \leq N_c$, since otherwise the Grid diagrams are no
longer convex.

Thus, we see that in a very simple way, the Grid diagrams and their
associated curves describe various types of flows in parameter
space. Some of these flows are easy to see also from the field theory
point of view, while others have no obvious field theory interpretation.


\begin{center}
\large{ACKNOWLEDGEMENTS}
\end{center}

BK would like to thank Lenny Susskind, his teacher; Arvind Rajaraman
for many hours of discussion; the Rutgers group for hospitality and
discussions and especially Tom Banks, Paul Aspinwall, Micha Berkooz,
Rami Entin, Shamit Kachru, Juan Maldacena, and Eva Silverstein; and
finally Renata Kallosh, Edi Halyo, Joachim Rahmfeld, Wing Kai Wong,
Jason Kumar and Peter Goldstein for their interest. OA would like to
thank Tom Banks, Jacques Distler, David Kutasov and Nathan Seiberg for useful
comments. AH would like to thank Cumrun Vafa for enlightening discussions.
BK is supported by NSF grant PHY-9219345. OA is supported in
part by DOE grant DE-FG02-96ER40559. AH is supported in part by NSF
grant PHY-9513835.

\bibliography{Grid}
\bibliographystyle{utphys}

\end{document}